\documentclass[12pt]{article}
\usepackage[latin9]{inputenc}
\usepackage{geometry}
\usepackage{amsmath}
\usepackage{amssymb}
\usepackage{stmaryrd}
\usepackage{graphicx}
\usepackage{setspace}
\usepackage[authoryear]{natbib}
\usepackage[unicode=true,
 bookmarks=false,
 breaklinks=false,pdfborder={0 0 1},backref=section,colorlinks=false]
 {hyperref}
\hypersetup{
 colorlinks,citecolor=blue,pdftex}

\makeatletter

\usepackage{amsfonts}
\usepackage{amsthm}
\usepackage{lscape}
\usepackage{appendix}
\usepackage{dcolumn}
\usepackage{rotating}
\usepackage{comment}
\usepackage{color}

\usepackage{pgf}
\usepackage{tikz}\usepackage{caption}
\usepackage{subcaption}
\usepackage{tkz-tab}
\usepackage{tikz-3dplot}
\usetikzlibrary{calc}
\usetikzlibrary{arrows, automata}

\newcolumntype{d}[1]{D{.}{.}{#1}}
\newcolumntype{t}[1]{D{,}{,}{#1}}
\newcolumntype{i}[1]{D{.}{}{#1}}
\newtheorem{theorem}{Theorem}[section]

\newtheorem{condition}{Condition}[section]

\newtheorem{corollary}{Corollary}[section]

\newtheorem{definition}{Definition}[section]
\newtheorem{example}{Example}
\newtheorem{lemma}{Lemma}[section]

\newtheorem{proposition}{Proposition}[section]
\newtheorem{remark}{Remark}[section]

\theoremstyle{plain} 

\newtheorem{innercustomthm}{Condition}
\newenvironment{mycondi}[1]
  {\renewcommand\theinnercustomthm{#1}\innercustomthm}
  {\endinnercustomthm}

\newtheorem{innercustomas}{Assumption}
\newenvironment{myas}[1]
  {\renewcommand\theinnercustomas{#1}\innercustomas}
  {\endinnercustomas}

\numberwithin{equation}{section}

\makeatother

\begin{document}
\title{On Quantile Treatment Effects, Rank Similarity,\\
and Variation of Instrumental Variables\thanks{For insightful discussions, the authors are grateful to Victor Chernozhukov
and participants at the 2023 North American Winter Meeting and the
Asian Meeting of the Econometric Society, the Barcelona School of
Economics Summer Forum, and the CeMMAP/SNU conference and in the seminars
at UCL, Oxford, BU and Boston College.}}
\author{Sukjin Han\\
 School of Economics\\
 University of Bristol\\
\UrlFont{\sffamily}\href{mailto:sukjin.han\%5C\%40gmail.com}{sukjin.han@gmail.com}\and
Haiqing Xu\\
Department of Economics \\
 University of Texas at Austin \\
\UrlFont{\sffamily}\href{mailto:h.xu\%5C\%40austin.utexas.edu}{h.xu@austin.utexas.edu}}
\date{\today}

\maketitle
\vspace{-0.2cm}

\begin{abstract}
This paper investigates how certain relationship between observed
and counterfactual distributions serves as an identifying condition
for treatment effects when the treatment is endogenous, and shows
that this condition holds in a range of nonparametric models for treatment
effects. To this end, we first provide a novel characterization of
the prevalent assumption restricting treatment heterogeneity in the
literature, namely rank similarity. Our characterization demonstrates
the stringency of this assumption and allows us to relax it in an
economically meaningful way, resulting in our identifying condition.
It also justifies the quest of richer exogenous variations in the
data (e.g., multi-valued or multiple instrumental variables) in exchange
for weaker identifying conditions. The primary goal of this investigation
is to provide empirical researchers with tools that are robust and
easy to implement but still yield tight policy evaluations.

\vspace{0.1in}


\end{abstract}

\section{Introduction\label{sec:Introduction}}

This paper investigates how certain relationship between observed
and counterfactual distributions serves as an identifying condition
for distributional treatment effects under endogeneity, and shows
that this condition holds in a range of nonparametric models for treatment
effects. To this end, we first provide a novel characterization of
the prevalent assumption restricting treatment heterogeneity in the
literature, namely rank similarity. Our characterization demonstrates
the stringency of this assumption and allows us to relax it in a economically
meaningful way, resulting in our identifying condition. It also justifies
the quest of richer exogenous variations in the data (e.g., multi-valued
or multiple instrumental variables) in exchange for the weaker identifying
condition.

The primary goal of this investigation is to provide empirical researchers
with (i) a framework where validity of identifying conditions prescribes
the parameters of interest, (ii) tools for identifying and estimating
treatment effects that allow for treatment heterogeneity, but that
still yield tight policy evaluation and are simple to implement, and
(iii) guidance on data collection that leads to drawing informative
causal conclusions.

Our analysis centers on the relationship between observed and counterfactual
distributions, specifically on the preservation of \enskip first-order
stochastic dominance (FOSD) of one distribution over the other to
their corresponding counterfactual distributions: for arbitrary compliance
types $t,t'\in\mathcal{T}$ induced by induced by individuals' potential
treatment responses to instrumental variables (IVs), if
\begin{align}
Y_{1}|t & \prec_{FOSD}Y_{1}|t'\label{eq:intro1}
\end{align}
then
\begin{align}
Y_{0}|t & \prec_{FOSD}Y_{0}|t',\label{eq:intro2}
\end{align}
where $Y_{d}$ denotes the counterfactual outcome given treatment
$D=d$.\footnote{For r.v.'s $A$ and $B$, let $A\prec_{FOSD}B$ denotes $F_{B}(\cdot)\le F_{A}(\cdot)$
where $F_{A}$ and $F_{B}$ are CDFs of $A$ and $B$, respectively.} This condition produces a partial ordering of the $Y_{0}$'s distributions
based on the partial ordering of the $Y_{1}$\textquoteright s distributions.
As we demonstrate later, this condition can be interpreted as $Y_{0}$
being ``noisier'' than $Y_{1}$ after controlling for all the confounding
variables.

We show that the proposed FOSD-preservation condition enables the
identification of certain counterfactual distributions, which are
essential components for identifying the treatment effects of our
interest. Only for the sake of illustration, consider \citet{imbens1994identification}'s
framework where binary instrument $Z\in\{0,1\}$ influences treatment
participation monotonically. Let compliance types $C$, $AT$, and
$NT$ stand for compliers, always-takers, and never-takers, respectively.
Let $Y$ be the observed outcome given by $Y\equiv DY_{1}+(1-D)Y_{0}$.
Suppose the observed distribution satisfies $Y_{1}|AT\prec_{FOSD}Y_{1}|C$.
Then the FOSD-preservation condition implies that $Y_{0}|AT\prec_{FOSD}Y_{0}|C$.
The latter provides an informative upper bound for $P[Y_{0}\le y|D=1]$
(and a symmetric analysis provides a lower bound), which is a necessary
component in calculating, for example, the quantile treatment effect
on the treated (QTT).\footnote{Note that $Y_{1}|AT\prec_{FOSD}Y_{1}|C$ and $Y_{0}|AT\prec_{FOSD}Y_{0}|C$
can be respectively rewritten as
\begin{align*}
P[Y\le\cdot|D=1,Z=0] & \le P[Y\le\cdot|D=1,Z=1],\\
P[Y_{0}\le\cdot|D=1] & \le\frac{P[Y\le\cdot,D=0|Z=1]-P[Y\le\cdot,D=0|Z=0]}{P[D=1|Z=0]-P[D=1|Z=1]}.
\end{align*}
}Although $Y_{1}|AT\prec_{FOSD}Y_{1}|C$ may seem restrictive, this
is \emph{not} generally the case when $Z$ departs from a scalar binary
variable. In this sense, our approach underscores the significance
of searching for richer exogenous variations of IVs, such as multi-valued
or multiple instrumental variables, as a means of trading for less
restrictive identifying conditions and achieving tighter bounds. Still,
the benefit of our approach can be manifested without requiring continuous
or large support of IVs. We also show that the proposed FOSD-preservation
condition (i.e., \eqref{eq:intro1} implies \eqref{eq:intro2}) yields
testable restrictions.

Nonparametric identification of treatment effects using IVs with limited
support has long been a challenging goal even when the focus is on
mean treatment effects, such as the average treatment effect (ATE)
and the ATE on the treated (ATT). In an influential line of literature,
\citet{manski1990nonparametric}, \citet{manski1997monotone}, and
\citet{MP00}, among many others, construct sharp bounds on the ATE
under a set of assumptions on the directions of treatment effects
and treatment selection while allowing instruments to be invalid in
a specific sense. Even with valid instruments, however, bounds on
the ATE are typically wide and uninformative to yield precise policy
prediction. The local ATE (LATE) (\citet{imbens1994identification})
and local QTE (\citet{abadie2002instrumental}) have been a popular
alternative when researchers are equipped with discrete IVs and impose
a monotonicity assumption on the selection to treatment. However,
the local group for which the treatment effect is identified may not
be the group of policy interest. Therefore, the extrapolation of the
local parameters becomes an important issue for policy analysis (e.g.,
treatment allocation), in which case the identification challenge
still remains (see e.g., \citet{mogstad2018using}, \citet{han2020sharp}).

Another prevalent approach in the literature is to restrict the degree
of treatment heterogeneity via rank similarity (or rank invariance).
This assumption has been shown to have substantial identifying power
for distributional treatment effects and the ATE and used in various
nonparametric contexts implicitly or explicitly (\citet{heckman1997making},
\citet{chesher2003identification,Che05}, \citet{chernozhukov2005iv},
\citet{VY07}, \citet{jun2011tighter}, \citet{SV11}, \citet{d2015identification},
\citet{torgovitsky2015identification}, \citet{vuong2017counterfactual},
\citet{han2021identification} to name a few). However, the plausibility
of this assumption can be questionable in many applications (e.g.,
\citet{maasoumi2019gender}) and testing methods are proposed as one
reaction to the skepticism (\citet{frandsen2018testing}, \citet{dong2018testing},
\citet{kim2022testing}).

In this paper, we clarify the stringency of the rank similarity assumption
by characterizing its restrictions on the relationship between observed
and counterfactual distributions. In particular, we show that the
\emph{strong} preservation of FOSD (i.e., \eqref{eq:intro1} holds
\emph{if and only if} \eqref{eq:intro2} holds) is equivalent to \emph{rank
linearity}, a slight relaxation of rank similarity that allows for
a linear transformation of an individual's rank to the counterfactual
rank. By doing so, we establish the connection between the rank similarity
assumption introduced by \citet{chernozhukov2005iv}'s structural
IV model and its corresponding conditions within \citet{rubin1974estimating}'s
counterfactual outcomes framework. Furthermore, we provide economic
justifications for the weak preservation of FOSD by proposing a variety
of non-separable structural IV models that imply the FOSD preservation
condition, but that do not satisfy rank similarity.

Based on our identification strategy, we develop a statistical linear
programming (LP) approach to estimate optimal bounds on the treatment
parameters. These bounds are defined as optimal values of LP (with
a discrete outcome) or semi-infinite LP (SILP) (with a continuous
outcome). To address the infeasibility of the SILP problem, we transform
the optimization problem by (i) randomizing the constraints or (ii)
invoking duality and approximates the Lagrangian measure using sieves.

The next section formally introduces the main identifying conditions
(i.e., the preservation of stochastic ordering) and establishes bounds
on treatment effects. Section \ref{sec:Justifying-Conditions-} introduces
structural models as sufficient conditions for the identifying conditions
presented in the previous section. Section \ref{sec:Systematic-Calculation-of}
discusses the computation of the bounds using linear programming and,
finally, Section \ref{sec:Numerical-Studies} presents numerical studies.
In the Appendix, Section \ref{sec:Point-Identification} shows that
point identification can be achieved with sufficient (but not infinite)
variation of IVs. In the main text we focus on the QTE. An extension
to bounding the ATE is discussed in Section \ref{sec:Conditions-for-Average}.
Section \ref{sec:Other-Structural-Models} contains more examples
of structural models as sufficient conditions and Section \ref{sec:Bounding-Violation-Probability}
holds further discussions on linear programming. All proofs are collected
in Section \ref{sec:Proofs}.

\section{Key Conditions and Bounds on Treatment Effects\label{sec:Setup-and-Key}}

Let $D\in\{0,1\}$ be the observed treatment indicator, which represents
the endogenous decision of an individual responding to IVs $Z$. We
assume $Z$ is either a vector of binary IVs or a multi-valued IV,
which takes $L$ distinct values: $Z\in\mathcal{Z}\equiv\{z_{1},...,z_{L}\}$.
Multi-valued or multiple IVs are common in many observational studies
(e.g., natural experiments typically provide more than one instrument)
and experimental studies (e.g., randomized control trials where multiple
treatment arms are implemented either simultaneously or sequentially).\footnote{See \citet{mogstad2021causal} for a recent survey.}
Let $Y_{1}$ be the counterfactual outcome of being treated and $Y_{0}$
be that of not being treated. They can be either continuously or discretely
distributed. The observed outcome $Y\in\mathcal{Y}\subseteq\mathbb{R}$
satisfies $Y=DY_{1}+(1-D)Y_{0}$. Finally, $X\in\mathcal{X}\subseteq\mathbb{R}^{k}$
denotes other covariates that may be endogenous.

Define QTE and ATE for treated and untreated populations. For $d\in\{0,1\}$
and $x\in\mathcal{X}$, define
\begin{align*}
QTE_{\tau}(d,x) & =Q_{Y_{1}|D,X}(\tau|d,x)-Q_{Y_{0}|D,X}(\tau|d,x)
\end{align*}
for $\tau\in(0,1)$ and
\begin{align*}
ATE(d,x) & =E[Y_{1}-Y_{0}|D=d,X=x].
\end{align*}
These parameters are what researchers and policymakers are potentially
interested. The unconditional QTE and ATE can be recovered when these
parameters are identified for all $d\in\{0,1\}$ and $x\in\mathcal{X}$.
Throughout the paper, we maintain that the IVs are valid and satisfy
the following exclusion restriction.

\begin{myas}{Z}\label{as:Z}For $d\in\{0,1\}$, $Z\perp Y_{d}|X$.\end{myas}

\subsection{Introducing Key Conditions}

Now we introduce the key condition that establishes the mapping between
observed and counterfactual distributions.

\begin{mycondi}{S$_{1}$}\label{as:condi_1}For arbitrary non-negative
weight vectors $(w_{1},...,w_{L})$ and $(\tilde{w}_{1},...,\tilde{w}_{L})$
that satisfy $\sum_{\ell=1}^{L}w_{\ell}=\sum_{\ell=1}^{L}\tilde{w}_{\ell}=1$,
if
\begin{align}
\sum_{\ell=1}^{L}w_{\ell}P[Y_{1}\le\cdot|D=1,Z=z_{\ell},X=x] & \le\sum_{\ell=1}^{L}\tilde{w}_{\ell}P[Y_{1}\le\cdot|D=1,Z=z_{\ell},X=x],\label{eq:condi1_0}
\end{align}
then
\begin{align}
\sum_{\ell=1}^{L}w_{\ell}P[Y_{0}\le\cdot|D=1,Z=z_{\ell},X=x] & \le\sum_{\ell=1}^{L}\tilde{w}_{\ell}P[Y_{0}\le\cdot|D=1,Z=z_{\ell},X=x].\label{eq:condi1_1}
\end{align}
\end{mycondi}

Importantly, note that in \eqref{eq:condi1_0} the probability $P[Y_{1}\le\cdot|D=1,Z=z_{\ell},X=x]$
can be obtained from the data as $Y_{1}=Y$ given $D=1$. The mapping
between observed and counterfactual distributions has been considered
in \citet{vuong2017counterfactual}, whose insights we share. Suppose
that $Z\perp(Y_{d},D_{z})|X$ additionally holds, where $D_{z}$ is
the counterfactual treatment given $Z=z$. Under this assumption,
each probability term in Condition \ref{as:condi_1} satisfies $P[Y_{d}\le\cdot|D=1,Z=z_{\ell},X=x]=P[Y_{d}\le\cdot|D_{z_{\ell}}=1,X=x]$.
Note that $\sum_{\ell=1}^{L}w_{\ell}P[Y_{1}\le\cdot|D_{z_{\ell}}=1,X=x]$
is a mixture of $Y_{d}$'s distributions weighted across different
compliance types defined by $\{D_{z_{\ell}}=1\}$, and thus can be
viewed as a distribution for a hypothetical population with the specific
composition of compliance types. Therefore, Condition \ref{as:condi_1}
posits that the FOSD ordering between the $Y_{1}$'s distributions
of two compliance compositions is preserved between their counterfactual
distributions of $Y_{0}$. For example, when $L=2$ and defiers are
excluded from possible compliance types (e.g., by \citet{imbens1994identification}'s
monotonicity assumption), then Condition \ref{as:condi_1} simply
describes the stochastic ordering between always-takers and compliers.
When $L\ge3$, however, there are more compliance types, which composition
becomes more complex as illustrated in Section \ref{sec:Interpretation-of-Ordering}.
Note that Condition \ref{as:condi_1} is \textit{not} an ``if and
only if'' statement. It would be stringent to impose the preservation
of ordering to hold in both directions. In fact, such a condition
is closely related to the rank similarity condition (\citet{chernozhukov2005iv});
see Section \ref{sec:Justifying-Conditions-} for full details. 

\subsection{Bounds on Treatment Effects}

Now, we show that Condition \ref{as:condi_1} is useful in constructing
bounds on $F_{Y_{0}|D,X}(\cdot|1,x)$ and subsequently on $QTE_{\tau}(1,x)$.
Let $p(z_{\ell},x)\equiv P[D=1|Z=z_{\ell},X=x]$ and
\begin{align*}
\Gamma(x) & \equiv\left\{ (\gamma_{1},...,\gamma_{L})\in\mathbb{R}^{L}:\sum_{\ell=1}^{L}\gamma_{\ell}=0\text{ and }\sum_{\ell=1}^{L}\gamma_{\ell}p(z_{\ell},x)=1\right\} .
\end{align*}
 \begin{theorem}\label{thm:main}Suppose that Assumption \ref{as:Z}
and Condition \ref{as:condi_1} hold. Fix $x\in\mathcal{X}$. For
$\gamma\equiv(\gamma_{1},...,\gamma_{L})$ and $\tilde{\gamma}\equiv(\tilde{\gamma}_{1},...,\tilde{\gamma}_{L})$
in $\Gamma(x)$, suppose 
\begin{align}
 & P[Y\leq\cdot|D=1,X=x]\leq\sum_{\ell=1}^{L}\gamma_{\ell}P[Y\leq\cdot,D=1|Z=z_{\ell},X=x],\label{eq:counter_k}\\
 & \sum_{\ell=1}^{L}\tilde{\gamma}_{\ell}P[Y\leq\cdot,D=1|Z=z_{\ell},X=x]\leq P[Y\leq\cdot|D=1,X=x].\label{eq:counter_k2}
\end{align}
Then $F_{Y_{0}|D,X}(\cdot|1,x)$ is bounded by
\begin{align}
 & -\sum_{\ell=1}^{L}\tilde{\gamma}_{\ell}P[Y\leq\cdot,D=0|Z=z_{\ell},X=x]\label{eq:LB}\\
 & \le P[Y_{0}\leq\cdot|D=1,X=x]\nonumber \\
 & \leq-\sum_{\ell=1}^{L}\gamma_{\ell}P[Y\leq\cdot,D=0|Z=z_{\ell},X=x]\label{eq:UB}
\end{align}
\end{theorem}

\begin{remark}[Constraints on $\gamma$]In Theorem \ref{thm:main},
$\Gamma(x)$ imposes two restrictions on $\gamma$: (i) $\sum_{\ell=1}^{L}\gamma_{\ell}=0$
and (ii) $\sum_{\ell=1}^{L}\gamma_{\ell}p(z_{\ell},x)=1$. First,
note that the existence of such a sequence requires the relevance
of the IV: $p(z_{\ell},x)\neq p(z_{\ell'},x)$ for some $z_{\ell},z_{\ell'}$.
Second, note that (ii) is a condition implied by either \eqref{eq:counter_k}
or \eqref{eq:counter_k2} with $y\rightarrow\infty$. Restriction
(ii) implicitly introduces a scale normalization. That is, for any
$\gamma$ satisfying $\sum_{\ell=1}^{L}\gamma_{\ell}p(z_{\ell},x)\neq0$,
we can always rescale it as $\gamma^{*}=\frac{\gamma}{\sum_{\ell=1}^{L}\gamma_{\ell}p(z_{\ell},x)}$
so that $\sum_{\ell=1}^{L}\gamma_{\ell}p(z_{\ell},x)=1$. It can be
shown that this normalization does not affect the bounds obtained
in \eqref{eq:LB} and \eqref{eq:UB}. \end{remark}

The proof of Theorem \ref{thm:main} and most of other proofs are
contained in the appendix. Note that there can be multiple $\gamma$
and $\tilde{\gamma}$ in $\Gamma(x)$ that satisfy \eqref{eq:counter_k}
and \eqref{eq:counter_k2}, respectively. Therefore, we can further
tighten the bounds as follows.

\begin{corollary}\label{corol:main}Suppose that Assumption \ref{as:Z}
and Condition \ref{as:condi_1} hold. Fix $x\in\mathcal{X}$. Then,
$F_{Y_{0}|D,X}(\cdot|1,x)$ is upper and lower bounded by
\begin{align*}
F_{Y_{0}|D,X}^{UB}(y|1,x) & \equiv\min_{\gamma\in\Gamma(x):\text{\eqref{eq:counter_k} holds}}-\sum_{\ell=1}^{L}\gamma_{\ell}P[Y\leq y,D=0|Z=z_{\ell}],\\
F_{Y_{0}|D,X}^{LB}(y|1,x) & \equiv\max_{\tilde{\gamma}\in\Gamma(x):\text{\eqref{eq:counter_k2} holds}}-\sum_{\ell=1}^{L}\tilde{\gamma}_{\ell}P[Y\leq y,D=0|Z=z_{\ell}].
\end{align*}

\end{corollary}

Theorem \ref{thm:main} and Corollary \ref{corol:main} highlight
the identifying power of multi-valued IVs. The key step in Theorem
\ref{thm:main} to calculate the bounds is to find $\gamma$ (resp.
$\tilde{\gamma}$) in $\Gamma(x)$ that satisfies \eqref{eq:counter_k}
(resp. \eqref{eq:counter_k2}), which serves as a rank condition.
Note that this condition is verifiable with the data. Corollary \ref{corol:main}
additionally implies that the bounds can be further tightened if one
increases the degree of freedom in the feasible set $\Gamma(x)$ by
increasing $L$, in which case \eqref{eq:counter_k}--\eqref{eq:counter_k2}
are more likely to hold. See below and Section \ref{sec:Numerical-Studies}
for related discussions.

Finally, note that
\[
QTE_{\tau}(1,x)=Q_{Y|D,X}(\tau|1,x)-Q_{Y_{0}|D,X}(\tau|1,x)
\]
and the bounds on the second quantity on the right-hand side can be
calculated using the worst case bounds for the conditional quantile
(\citet{manski1994selection}, \citet{blundell2007changes}):
\begin{align*}
Q_{Y_{0}|D,X}^{LB}(\tau|1,x) & \le Q_{Y_{0}|D,X}(\tau|1,x)\le Q_{Y_{0}|D,X}^{UB}(\tau|1,x),
\end{align*}
where $Q_{Y_{0}|D,X}^{LB}(\tau|1,x)$ and $Q_{Y_{0}|D,X}^{UB}(\tau|1,x)$
are the $\tau$-th quantiles of $F_{Y_{0}|D,X}^{LB}(\cdot|1,x)$ and
$F_{Y_{0}|D,X}^{UB}(\cdot|1,x)$, respectively. Although the bounds
on $ATE(1,x)=E[Y|D=1,X=x]-E[Y_{0}|D=1,X=x]$ can be calculated based
on $E[Y_{0}|D=1,X=x]=\int_{0}^{1}Q_{Y_{0}|D,X}(\tau|1,x)d\tau$, we
present later how the bounds on the $ATE(d,x)$ can be calculated
under a weaker condition than Condition \ref{as:condi_1}.

If we assume the converse of Condition \ref{as:condi_1}, we can calculate
bounds on the $QTE(0,x)$.

\begin{mycondi}{S$_{0}$}\label{as:condi_0}For arbitrary non-negative
weight vectors $(w_{1},...,w_{L})$ and $(\tilde{w}_{1},...,\tilde{w}_{L})$
that satisfy $\sum_{\ell=1}^{L}w_{\ell}=\sum_{\ell=1}^{L}\tilde{w}_{\ell}=1$,
if
\begin{align}
\sum_{\ell=1}^{L}w_{\ell}P[Y_{0}\le\cdot|D=0,Z=z_{\ell},X=x] & \le\sum_{\ell=1}^{L}\tilde{w}_{\ell}P[Y_{0}\le\cdot|D=0,Z=z_{\ell},X=x],\label{eq:condi0_0}
\end{align}
then
\begin{align}
\sum_{\ell=1}^{L}w_{\ell}P[Y_{1}\le\cdot|D=0,Z=z_{\ell},X=x] & \le\sum_{\ell=1}^{L}\tilde{w}_{\ell}P[Y_{1}\le\cdot|D=0,Z=z_{\ell},X=x].\label{eq:condi0_1}
\end{align}
\end{mycondi}

Similar to Condition \ref{as:condi_1}, Condition \ref{as:condi_0}
ensures that if the two distributions of $Y_{0}$ exhibit an FOSD
ordering across different compliance compositions, this ordering is
preserved in their corresponding counterfactual $Y_{1}$ distributions.
In practice, depending on the specific context, Condition \ref{as:condi_1},
Condition \ref{as:condi_0}, or both may hold; Section \ref{sec:Justifying-Conditions-}
provides some related intuitions. Not surprisingly, we can use Condition
\ref{as:condi_0} to construct bounds on $F_{Y_{1}|D,X}(\cdot|0,x)$
and subsequently on $QTE_{\tau}(0,x)$.

\begin{theorem}\label{thm:main_0}Suppose that Assumption \ref{as:Z}
and Condition \ref{as:condi_0} hold. Fix $x\in\mathcal{X}$. For
$\gamma\equiv(\gamma_{1},...,\gamma_{L})$ and $\tilde{\gamma}\equiv(\tilde{\gamma}_{1},...,\tilde{\gamma}_{L})$
in $\Gamma(x)$, suppose 
\begin{align}
 & P[Y\leq\cdot|D=0,X=x]\leq\sum_{\ell=1}^{L}\gamma_{\ell}P[Y\leq\cdot,D=0|Z=z_{\ell},X=x],\label{eq:counter_k-0}\\
 & \sum_{\ell=1}^{L}\tilde{\gamma}_{\ell}P[Y\leq\cdot,D=0|Z=z_{\ell},X=x]\leq P[Y\leq\cdot|D=0,X=x].\label{eq:counter_k2-0}
\end{align}
Then $F_{Y_{1}|D,X}(\cdot|0,x)$ is bounded by
\begin{align}
 & -\sum_{\ell=1}^{L}\tilde{\gamma}_{\ell}P[Y\leq\cdot,D=1|Z=z_{\ell},X=x]\label{eq:LB-0}\\
 & \le P[Y_{1}\leq\cdot|D=0,X=x]\nonumber \\
 & \leq-\sum_{\ell=1}^{L}\gamma_{\ell}P[Y\leq\cdot,D=1|Z=z_{\ell},X=x].\label{eq:UB-0}
\end{align}
\end{theorem}

The proof of this theorem is analogous to that of Theorem \ref{thm:main}.
The bounds on $QTE_{\tau}(0,x)$ can be derived symmetrically as in
the case of $QTE_{\tau}(1,x)$ and thus are omitted. Notably, which
treatment parameter we can obtain bounds for is determined by which
identifying condition we impose (i.e., Condition \ref{as:condi_1}
or \ref{as:condi_0}). In Section \ref{sec:Justifying-Conditions-},
we investigate this aspect within economic structural models. Finally,
in the Appendix, we introduce weaker conditions to bound average treatment
effects.

\subsection{Understanding Key Conditions\label{sec:Interpretation-of-Ordering}}

We further explore Conditions \ref{as:condi_1} and \ref{as:condi_0}
to give additional interpretation and discuss testability. Suppress
$X=x$ to simplify our discussions. Under $Z\perp(Y_{d},D_{z})$,
the inequalities for FOSD in Conditions \ref{as:condi_1} and \ref{as:condi_0}
can be rewritten as
\begin{align}
\sum_{\ell=1}^{L}w_{\ell}P[Y_{d}\le y|D_{z_{\ell}}=1] & \le\sum_{\ell=1}^{L}\tilde{w}_{\ell}P[Y_{d}\le y|D_{z_{\ell}}=1].\label{eq:condi1_2}
\end{align}
Recall, Theorem \ref{thm:main} relies on the existence of a sequence
$\gamma=(\gamma_{1},...,\gamma_{L})$ satisfying $\sum_{\ell=1}^{L}\gamma_{\ell}=0$,
$\sum_{\ell=1}^{L}\gamma_{\ell}p(z_{\ell},x)=1$, and the inequality
\eqref{eq:counter_k}, that is, $P[Y\leq y|D=1]\leq\sum_{\ell=1}^{L}\gamma_{\ell}P[Y\leq y,D=1|Z=z_{\ell}]$
for all $y$. Note that \eqref{eq:counter_k} is a special case of
\eqref{eq:condi1_2} with $d=1$, which is the ``if'' part of Condition
\ref{as:condi_1}. Let $p(z)\equiv(D=1|Z=z)$. Only for the purpose
of this subsection, assume the generalized version of the LATE monotonicity
introduced in \citet{imbens1994identification}:
\begin{align}
\text{For }\ell\neq\ell', & \text{ either }D_{z_{\ell}}\ge D_{z_{\ell'}}\text{ a.s. or }D_{z_{\ell}}\le D_{z_{\ell'}}\text{ a.s.}\label{eq:as_D*}
\end{align}

Under \eqref{eq:as_D*}, $\{D_{z_{\ell}}=1\}$ in \eqref{eq:condi1_2}
are a mix of individuals who are compliers (C) and always-takers (AT).
For $Z\in\mathcal{Z}=\{z_{1},...,z_{L}\}$, let $(z_{\ell-1},z_{\ell})$-compliers
be compliers induced by the change of $Z$ from $z_{\ell-1}$ to $z_{\ell}$.
When $L=3$ and $(z_{1},z_{2},z_{3})=(0,1,2)$, for example, $\{(0,1)\text{-C}\}=\{i:D_{0,i}=0,D_{1,i}=D_{2,i}=1\}$
is the set of eager compliers (E-C) and $\{(1,2)\text{-C}\}=\{i:D_{0,i}=D_{1,i}=0,D_{2,i}=1\}$
is the set of reluctant compliers (R-C), following the language of
\citet{mogstad2021causal}. Also, $\{\text{AT}\}=\{i:D_{0,i}=D_{1,i}=D_{2,i}=1\}$
is the set of always-takers. Let $p_{\ell}$ for $\ell=\{2,...,L\}$
is the proportion of $(z_{\ell-1},z_{\ell})$-compliers and let $p_{1}\equiv P[\text{AT}]$.
We show that \eqref{eq:condi1_2} establishes the FOSD relationship
between the mixtures of observed distributions of $Y$ conditional
on various always-takers and compliers groups:

\begin{lemma}\label{lem:refutable}Suppose \eqref{eq:as_D*} holds
and $Z\perp(Y_{d},D_{z})$ and $0<p(z_{\ell})<1$ for all $\ell$.
(i) Then, \eqref{eq:condi1_2} is equivalent to
\begin{align*}
 & \omega_{1}P[Y_{d}\le y|\text{AT}]+\sum_{\ell=2}^{L}\omega_{\ell}P[Y_{d}\le y|(z_{\ell-1},z_{\ell})\text{-C}]\\
 & \le\tilde{\omega}_{1}P[Y_{d}\le y|\text{AT}]+\sum_{\ell=2}^{L}\tilde{\omega}_{\ell}P[Y_{d}\le y|(z_{\ell-1},z_{\ell})\text{-C}]
\end{align*}
for some non-negative $\omega_{\ell}$ and $\tilde{\omega}_{\ell}$
for $\ell=1,...,L$. (ii) Moreover, suppose $L\ge2$ and $w_{1}+p_{1}\sum_{\ell=2}^{L}\frac{w_{\ell}}{p(z_{\ell})}=\tilde{w}_{1}+p_{1}\sum_{\ell=2}^{L}\frac{\tilde{w}_{\ell}}{p(z_{\ell})}$.
Then \eqref{eq:condi1_2} with $w\neq\tilde{w}$ can be expressed
as
\begin{align}
\sum_{\ell=2}^{L}\omega_{\ell}P[Y_{d}\le y|(z_{\ell-1},z_{\ell})\text{-C}] & \le\sum_{\ell=2}^{L}\tilde{\omega}_{\ell}P[Y_{d}\le y|(z_{\ell-1},z_{\ell})\text{-C}]\label{eq:refutable}
\end{align}
for some non-negative $\omega_{\ell}$ and $\tilde{\omega}_{\ell}$
for $\ell=2,...,L$.\end{lemma}

To illustrate the intuition of Lemma \ref{lem:refutable}(i), consider
$L=3$ and $(z_{1},z_{2},z_{3})=(0,1,2)$. Then, 
\begin{align*}
\{D_{1}=1\} & =\{D_{0}=1,D_{1}=1,D_{2}=1\}\cup\{D_{0}=0,D_{1}=1,D_{2}=1\}=\{\text{AT}\}\cup\{\text{E-C}\},
\end{align*}
because $\{D_{0}=1,D_{1}=1,D_{2}=0\}=\emptyset$ and $\{D_{0}=0,D_{1}=1,D_{2}=0\}=\emptyset$.
Also, $\{D_{2}=1\}=\{\text{AT}\}\cup\{\text{E-C}\}\cup\{\text{R-C}\}$
and $\{D_{0}=1\}=\{\text{AT}\}$. 

Lemma \ref{lem:refutable}(ii) can be used as the basis to test \eqref{eq:condi1_2}
and thus Condition \ref{as:condi_1}. The intuition is as follows.
With a binary IV, the marginal distributions of $Y_{1}$ and $Y_{0}$
are identified for compliers (\citet{abadie2002instrumental}). This
result holds for any complier group defined by a pair of instrument
values, such as $\{(z_{\ell-1},z_{\ell})\text{-C}\}$ in the lemma.
Then, when $L\ge2$, we can find vectors $w$ and $\tilde{w}$ in
$\mathbb{R}_{+}^{L}$ that assign zero weights to the distributions
for AT and still make \eqref{eq:condi1_2} a non-trivial inequality
where all the associated distributions for compliers are identified
for all $d=1,0$.

\begin{remark}[Conditions w.r.t. Compliance Types]\label{rem:compliance_type}Motivated
from the discussion of this section, we can rewrite Condition \ref{as:condi_1}
(and all the relevant conditions) solely in terms of compliance types.
Let $T\equiv\{D(z_{1}),...,D(z_{L})\}$ be a random vector that indicates
a particular compliance type with its realized value in $\{0,1\}^{L}\equiv\tilde{\mathcal{T}}$.
For example, when $L=2$ (i.e., binary IV), $T\equiv(D(0),D(1))\in\{(0,0),(1,0),(0,1),(1,1)\}\equiv\tilde{\mathcal{T}}$.
Since $D$ and $Z$ are discrete, $T$ is naturally a discrete random
vector. Note that this framework do not rely on any selection models,
and therefore $T$ captures all possible compliance types given $D$
and $Z$. Then Condition \ref{as:condi_1} can be rewritten into the
following slightly stringent one:

\emph{Fix $x\in\mathcal{X}$. For arbitrary weight functions $w:\tilde{\mathcal{T}}\times\mathcal{X}\rightarrow\mathbb{R}_{+}$
and $\tilde{w}:\tilde{\mathcal{T}}\times\mathcal{X}\rightarrow\mathbb{R}_{+}$
such that $\sum_{t\in\tilde{\mathcal{T}}}w(t,x)=\sum_{t\in\tilde{\mathcal{T}}}\tilde{w}(t,x)=1$,
if 
\[
\sum_{t\in\tilde{\mathcal{T}}}w(t,x)F_{Y_{1}|T,X}(\cdot|t,x)\leq\sum_{t\in\tilde{\mathcal{T}}}\tilde{w}(t,x)F_{Y_{1}|T,X}(\cdot|t,x),
\]
then
\[
\sum_{t\in\tilde{\mathcal{T}}}w(t,x)F_{Y_{0}|T,X}(\cdot|t,x)\leq\sum_{t\in\tilde{\mathcal{T}}}\tilde{w}(t,x)F_{Y_{0}|T,X}(\cdot|t,x).
\]
}

Then, the weighted sum in each inequality can be interpreted as the
distribution of $Y_{d}$ weighted across all compliance types.

\end{remark}

\section{Structural Models as Sufficient Conditions\label{sec:Justifying-Conditions-}}

We show that Conditions \ref{as:condi_1} and \ref{as:condi_0} can
be justified in a range of nonparametric structural models for the
counterfactual outcomes. To this end, it is useful to first present
a stronger version of Condition \ref{as:condi_1} (labeled as \ref{condi_1_suff}).
This version of the condition is motivated by the discussion in Remark
\ref{rem:compliance_type}. To state this condition, we introduce
a general model for treatment selection:

\begin{myas}{D}\label{as:D}Assume that
\begin{align}
D & =h(Z,X,\eta),\label{eq:sel}
\end{align}
where $\eta\in\mathcal{T}$ can be an arbitrary vector.\end{myas}

Note that Assumption \ref{as:D} permits a more general compliance
behavior than what a weakly separable model $D=1\{\eta\le h(Z,X)\}$
does (or equivalently, \eqref{eq:as_D*} as shown in \citet{vytlacil2002independence}).
Although Assumption \ref{as:D} is not necessary for our main procedure,
it is useful in defining the types of compliance behavior (i.e., treatment
selection mechanism) via the unobservable $\eta$. Under this assumption,
the following condition implies Condition \ref{as:condi_1}.\footnote{More precisely, it implies the condition in Remark \ref{rem:compliance_type},
which in turn implies Condition \ref{as:condi_1}.} Let $F_{Y_{d}|\eta,X}(y|t,x)\equiv P[Y_{d}\le y|\eta=t,X=x]$.

\begin{mycondi}{S$_{1}^{*}$}\label{condi_1_suff}Fix $x\in\mathcal{X}$.
For arbitrary weight functions $w:\mathcal{T}\times\mathcal{X}\rightarrow\mathbb{R}_{+}$
and $\tilde{w}:\mathcal{T}\times\mathcal{X}\rightarrow\mathbb{R}_{+}$
such that $\int w(t,x)dt=\int\tilde{w}(t,x)dt=1$, if 
\[
\int w(t,x)F_{Y_{1}|\eta,X}(\cdot|t,x)dt\leq\int\tilde{w}(t,x)F_{Y_{1}|\eta,X}(\cdot|t,x)dt,
\]
then
\[
\int w(t,x)F_{Y_{0}|\eta,X}(\cdot|t,x)dt\leq\int\tilde{w}(t,x)F_{Y_{0}|\eta,X}(\cdot|t,x)dt.
\]
\end{mycondi}

Because $w(\cdot,x)$ is non-negative and $\int w(t,x)dt=1$, note
that $\int w(t,x)F_{Y_{d}|\eta,X}(\cdot|t,x)dt$ is a mixture of conditional
CDFs (with $w(\cdot,x)$ being the mixture weight) and thus itself
a CDF. In other words, defining a type distribution $W_{x}(t)=\int^{t}w(\eta,x)d\eta$,
we can write $\int w(t,x)F_{Y_{d}|\eta,X}(\cdot|t,x)dt=\int F_{Y_{d}|\eta,X}(\cdot|t,x)dW_{x}(t)$.\footnote{Since $\eta$ has arbitrary dimensions, the integral with respect
to $t$ is understood to be a multivariate integral.} Therefore, Condition \ref{condi_1_suff} assumes that the FOSD ordering
of $Y_{1}$ distributions conditional on $\eta$ conforming to two
different type distributions ($W_{x}(\cdot)$ and $\tilde{W}_{x}(\cdot)$)
is preserved in the ordering of $Y_{0}$ distributions conditional
on the corresponding type distributions.

The following lemma establishes the sufficiency of Condition \ref{condi_1_suff}
for Condition \ref{as:condi_1}.

\begin{lemma}\label{lem:suff}Under Assumption \ref{as:D}, Condition
\ref{condi_1_suff} implies Condition \ref{as:condi_1}.\end{lemma}

Symmetrically, Condition \ref{as:condi_0} has a corresponding stronger
condition, which is omitted.

Now, we relate the conditions with the structural models, which provide
additional intuitions for the conditions. We present a leading model
here and the rest in the Appendix. For arbitrary r.v.'s $A$ and $\tilde{A}$,
let $A\stackrel{d}{=}\tilde{A}$ denote $F_{A}=F_{\tilde{A}}$.

\bigskip{}

\noindent \textbf{Model 1.} (i) Assumption \ref{as:D} holds and
\begin{align}
Y & =q(D,X,U_{D}),\label{eq:model1-1}
\end{align}
where $q(d,x,\cdot)$ is continuous and monotone increasing and $U_{D}=DU_{1}+(1-D)U_{0}$,
(ii) conditional on $(\eta,X,Z)$, $U_{d}\stackrel{d}{=}U+\xi_{d}$
where $\xi_{d}\perp(\eta,U)$, (iii) conditional on $(X,Z)$, $\xi_{0}$
is (weakly) more or less noisy than $\xi_{1}$, that is, $\xi_{0}\stackrel{d}{=}\xi_{1}+V$
for some $V$ independent of $\xi_{1}$.\bigskip{}

Note that $U$ is the source of endogeneity in that it allowed to
be dependent on $\eta$. Model 1(ii)--(iii) implies that $U_{0}\stackrel{d}{=}U_{1}+V$
conditional on $(\eta,X,Z)$. Importantly, Model 1 nests the model
in \citet{chernozhukov2005iv} as a special case. This can be shown
as follows. First, they assume Model 1(i) and, conditional on $(X,Z)$,
either rank similarity ($F_{U_{0}|\eta}=F_{U_{1}|\eta}$) or rank
invariance ($U_{0}=U_{1}$).\footnote{Note that rank similarity and rank invariance are observationally
equivalent under Model 1(i) in that they produce the same distribution
of observables (\citet{chernozhukov2013quantile}).} Then, by taking $\xi_{d}=0$ for all $d$ in Model 1(ii), we have
$U_{0}\stackrel{d}{=}U_{1}\stackrel{d}{=}U$ conditional on $(\eta,X,Z)$,
which proves the claim.

Model 1(iii) assumes that the unobservable under the counterfactual
status of being treated are more (or less) dispersed than that under
the counterfactually untreated status. Although this may seem stringent,
it is substantially weaker than rank similarity (or invariance) and
can be plausible in various scenarios. Before providing examples of
these scenarios, we first establish the connection between Model 1
and Condition \ref{condi_1_suff} (and thus Condition \ref{as:condi_1}
by Lemma \ref{lem:suff}).

\begin{theorem}\label{thm:model1}Under Assumptions Z, Model 1 (with
$\xi_{0}$ being weakly more noisy than $\xi_{1}$) implies Condition
\ref{condi_1_suff} and thus Condition \ref{as:condi_1}.\end{theorem}Analogous
to Theorem \ref{thm:model1}, one can readily show that Model 1 with
$\xi_{0}$ being weakly less noisy than $\xi_{1}$ implies Condition
\ref{as:condi_0}.

Now we provide examples that are consistent with Model 1.

\begin{example}[Auction]\label{ex1}Consider online and offline auctions.
Let $Y$ be the bid (which subsequently forms revenue) and $D$ be
participating in an auction with different format ($D=1$ if online
and $=0$ if offline). Let $U_{d}\stackrel{d}{=}U+\xi_{d}$ be the
valuation of the item where $U$ is the common valuation (correlated
with $D$) and $\xi_{d}$ is format specific random shocks satisfying
$\xi_{d}\perp(\eta,U)$. We assume that bidders have limited information
on certain features of the auction that affect valuation (e.g., they
know the distribution of $\xi_{d}$ but not its realization). In this
example, what would justify $var(\xi_{0})>var(\xi_{1})$? It may be
the case that, in the offline auction, bidders are more emotionally
affected by other bidders, which makes their bids more variable.\end{example}

\begin{example}[Insurance]\label{ex2}We are interested in the effect
of insurance on health outcomes. Let $Y$ be the health outcome and
$D$ be the decision of getting insurance ($D=1$ being insured).
Let $U_{d}\stackrel{d}{=}U+\xi_{d}$ be underlying health conditions
where $U$ captures health conditions known to participant (and thus
correlated with $D$) while $\xi_{d}$ is health conditions not fully
known a priori and thus random. In this example, $var(\xi_{0})>var(\xi_{1})$
may hold because insurance by definition ensures a certain level of
health conditions.\end{example}

\begin{example}[Vaccination]\label{ex3}Similar to Example \ref{ex2},
suppose that $D$ is instead getting vaccination (of an established
vaccine). Again, $U_{d}\stackrel{d}{=}U+\xi_{d}$ is health conditions
where $U$ captures conditions known to participant (and correlated
with $D$) and $\xi_{d}$ is vaccination-status-specific health conditions,
which are not fully known a priori. Then, similarly as before $var(\xi_{0})>var(\xi_{1})$
may hold because, when not vaccinated, one is exposed to the risk
of a serious illness, while vaccination ensures a certain level of
immunity.\end{example}

The scenarios in Examples \ref{ex1}--\ref{ex3} justify Condition
\ref{as:condi_1} via Theorem \ref{thm:model1}. Then, under Condition
\ref{as:condi_1}, Theorem \ref{thm:main} and Corollary \ref{corol:main}
yield bounds on $QTE_{\tau}(1,x)$, the effects of treatment for those
who take the treatment. The final example illustrates the converse
case.

\begin{example}[Medical Trial]\label{ex4}In contrast to Example
\ref{ex3}, suppose the treatment itself is risky. That is, let $D$
be participating in a frontier medical trial ($D=1$ being participation).
In this case, $var(\xi_{0})<var(\xi_{1})$ is more plausible because,
with a newly developed medicine, there is the high risk of unknown
side effects.\end{example}

The scenario in Example \ref{ex4} justifies Condition \ref{as:condi_0},
under which bounds on $QTE_{\tau}(0,x)$, the effects of treatment
for those who abstain from it, can be obtained.

Model 1 and these examples show how a certain treatment parameter
may be more relevant for policy than others depending on the plausibility
of assumptions. Consider the problem of a policymaker. Assume that
the policymaker concerns risk-averse individuals, \emph{which are
typically the majority}. For this policymaker, a candidate policy
would aim at providing ``insurance,'' which can be either literally
insurance or policies that serve as insurance (e.g., vaccination,
subsidies). Therefore, she would be interested in understanding the
treatment effects for the target individuals that are risk-averse.
Our procedure provides a statistical tool for such a policymaker.
That is, under Model 1, our procedure has the ability to bound the
treatment effects for individuals with $D=d$ such that $var(\xi_{d})<var(\xi_{1-d})$.
This is a unique feature of our setting: the plausibility of assumptions
dictates the parameters of interest, which then can be terms as \emph{assumption-driven}
treatment parameters.

A remaining question one might have is as follows. How much Condition
\ref{as:condi_1} has to be strengthened to be equivalent to rank
similarity? To answer this question, recall that Condition \ref{condi_1_suff}
is stronger than Condition \ref{as:condi_1} (by Lemma \ref{lem:suff}).
We strengthen Condition \ref{condi_1_suff} further by making it an
``if and only if'' condition:

\begin{mycondi}{S$^{*}$}\label{condi_2}Fix $x\in\mathcal{X}$. For
arbitrary weight functions $w:\mathcal{T}\times\mathcal{X}\rightarrow\mathbb{R}_{+}$
and $\tilde{w}:\mathcal{T}\times\mathcal{X}\rightarrow\mathbb{R}_{+}$
such that $\int w(t,x)dt=\int\tilde{w}(t,x)=1$, it holds that
\[
\int w(t,x)F_{Y_{1}|\eta,X}(\cdot|t,x)dt\leq\int\tilde{w}(t,x)F_{Y_{1}|\eta,X}(\cdot|t,x)dt
\]
if and only if
\[
\int w(t,x)F_{Y_{0}|\eta,X}(\cdot|t,x)dt\leq\int\tilde{w}(t,x)F_{Y_{0}|\eta,X}(\cdot|t,x)dt.
\]
\end{mycondi}

It turns out that we can establish the following result.

\begin{theorem}\label{thm:model2}Model 1(i) with $F_{U_{0}|\eta,X,Z}=F_{U_{1}|\eta,X,Z}$
(i.e., rank similarity) implies Condition \ref{condi_2}.\end{theorem}

This theorem highlights the stringency of rank similarity relative
to Condition \ref{as:condi_1}, which is used in our bound analysis.
The proof is trivial so omitted. It is worth noting that the converse
of Theorem \ref{thm:model2} is \emph{not} true. Here is a counter-example
for the converse statement.

\begin{definition}[Rank Linearity]Assume Model 1(i) and
\begin{align}
F_{Y_{0}|\eta,X,Z}(\cdot|t,x,z) & =\lambda(\cdot,x)F_{Y_{1}|\eta,X,Z}(\psi(\cdot,x)|t,x,z)\label{eq:rank_linearity}
\end{align}
for every $t\in\mathcal{T}$, $x\in\mathcal{X}$ and $z\in\mathcal{Z}$,
where $\psi(\cdot,x):\mathcal{Y}\rightarrow\mathcal{Y}$, a one-to-one
and onto mapping, is strictly increasing, and $\lambda(\cdot,x):\mathcal{Y}\rightarrow\mathbb{R}_{+}$
is consistent with $F_{Y_{d}|\eta,X,Z}$ being a proper CDF.\end{definition}

This \emph{rank linearity} implies Condition \ref{condi_2}, which
is trivial to show. However, rank linearity is weaker than rank similarity
as the latter is a special case of the former. To see this, conditional
on $Z=z$ (and suppressing $X)$, \eqref{eq:rank_linearity} with
Model 1(i) yields $F_{U_{0}|\eta}(q^{-1}(0,y)|t)=\lambda(y)F_{U_{1}|\eta}(q^{-1}(1,\psi(y))|t)$.
Then, by choosing $\lambda(y)=1$ and $\psi(y)=\phi(y)\equiv q(1,q^{-1}(0,y))$
(i.e., the counterfactual mapping (\citet{vuong2017counterfactual})),
we have $F_{U_{0}|\eta}(\cdot|t)=F_{U_{1}|\eta}(\cdot|t)$.\footnote{Alternatively, rank similarity can be equivalently stated as $F_{Y_{0}|\eta}(y|t)=F_{Y_{1}|\eta}(\phi(y)|t)$
(where $\phi(y)$ is strictly increasing), which can be derived from
\eqref{eq:rank_linearity} by choosing $\lambda(y)=1$ and $\psi(y)=\phi(y)$.} In general, while the ranks between $Y_{0}$ and $Y_{1}$ should
have the same distribution under rank similarity, their distributions
can be different under rank linearity because of the multiplying
term $\lambda(\cdot)$ in $F_{U_{0}|\eta}(u|t)=\lambda(q(0,u))F_{U_{1}|\eta}(u|t)$.
However, note that the the difference cannot be entirely arbitrary
as $\lambda(\cdot)$ does not depend on $t$, and thus rank linearity
still poses substantive restrictions.

Interestingly, rank linearity is equivalent to Condition \ref{condi_2}.
The following theorem is one of the main contributions of this paper.
Suppress $(Z,X)$ for simplicity.

\begin{theorem}\label{thm:equiv_RL}Suppose for any CDF $F_{1}(\cdot)$
supported on $\mathbb{R}$, there always exists a function $c:\mathcal{T}\rightarrow\mathbb{R}$
such that 
\begin{equation}
F_{d}(\cdot)=\int c(t)F_{Y_{d}|\eta}(\cdot|t)dt.\label{eq:rank_condi}
\end{equation}
Then Condition \ref{condi_2} holds if and only if there exits some
$\psi(\cdot)$ that is strictly increasing and $\lambda(\cdot)>0$
such that 
\begin{equation}
F_{Y_{0}|\eta}(\cdot|t)=\lambda(\cdot)F_{Y_{1}|\eta}(\psi(\cdot)|t)\qquad\text{for }t\in\mathcal{T}.\label{eq:RL-1}
\end{equation}

\end{theorem}

We prove this equivalence in the Appendix. The proof with continuous
$Y_{d}$ is more involved than that with discrete $Y_{d}$; we recommend
that the interested reader reads the latter first. The condition \eqref{eq:rank_condi}
is only introduced in this theorem to establish the relationship between
rank linearity (and hence rank similarity) and the range of identifying
conditions of this paper, and it is not necessary for our bound analysis.
This condition would be violated when there is no endogeneity (i.e.,
$Y_{d}\perp\eta$), which is not our focus.

Condition \ref{condi_2} is crucial in bounding $QTE_{\tau}(x)=Q_{Y_{1}|X}(\tau|x)-Q_{Y_{0}|X}(\tau|x)$
unconditional with respect to $D=d$. The ``only if'' part (i.e.,
Condition \ref{as:condi_1}) will bound $Q_{Y_{0}|D=1}(\tau)$ and
thus $Q_{Y_{0}}(\tau)$ by Theorem \ref{thm:main}, while the ``if''
part (i.e., Condition \ref{as:condi_0}) will bound $Q_{Y_{1}|D=0}(\tau)$
and thus $Q_{Y_{1}}(\tau)$ by the symmetric version of Theorem \ref{thm:main}.
The fact that Condition \ref{condi_2} is weaker than rank similarity
illustrates the importance of rank similarity in the identification
of the QTE and ATE.

\begin{remark}[Testability of the Conditions]It is immediate from
Lemma \ref{lem:refutable}(ii) that Condition \ref{condi_2} can be
tested from the data when $L\ge2$ and under the LATE monotonicity
assumption. Given the established connection between Condition \ref{condi_2}
and rank similarity (Theorem \ref{thm:model2}), when Condition \ref{condi_2}
is refuted from the data, rank similarity can be refuted. This result
relates to the testability of rank similarity under LATE monotonicity
(\citet{dong2018testing}, \citet{kim2022testing}).

\end{remark}

\begin{remark}[Conditions w.r.t. Compliance Types, continued]Continuing
the discussion in Remark \ref{rem:compliance_type}, $F_{U_{1}|T,Z}=F_{U_{0}|T,Z}$
(where $T\equiv(D(0),D(1))$ and $X$ is suppressed) can be viewed
as an alternative rank similarity assumption. Because $\sigma(T)\subset\sigma(\eta)$
where $\sigma(A)$ is a $\sigma$-field generated by a random vector
$A$, $F_{U_{1}|\eta,Z}=F_{U_{0}|\eta,Z}$ implies $F_{U_{1}|T,Z}=F_{U_{0}|T,Z}$.
Then \citet{chernozhukov2005iv}'s main testable implication ((2.6)
in Theorem 1 of their paper) can be equally derived under $F_{U_{1}|T,Z}=F_{U_{0}|T,Z}$,
which clarifies the role of selection mechanism in their analysis.
To see this, let $t\equiv(t_{0},t_{1})$ be the realization of $T\equiv(D(0),D(1))$
and assume Model 1(i). We have
\begin{align*}
P[Y\le q(D,\tau)|Z=z] & =P[q(D,U_{D})\le q(D,\tau)|Z=z]\\
 & =P[U_{D}\le\tau|Z=z]\\
 & =\sum_{t\in\tilde{\mathcal{T}}}P[U_{D(z)}\le\tau|Z=z,T=t]P[T=t|Z=z]
\end{align*}
but
\begin{align*}
\sum_{t\in\tilde{\mathcal{T}}}P[U_{D(z)}\le\tau|Z=z,T=t]P[T=t|Z=z] & =\sum_{t\in\tilde{\mathcal{T}}}P[U_{t_{z}}\le\tau|Z=z,T=t]P[T=t|Z=z]\\
 & =\sum_{t\in\tilde{\mathcal{T}}}P[U_{0}\le\tau|Z=z,T=t]P[T=t|Z=z]\\
 & =P[U_{0}\le\tau|Z=z]\\
 & =\tau,
\end{align*}
where $F_{U_{1}|T,Z}=F_{U_{0}|T,Z}$ is used in the second equality
and $U_{0}\perp Z$ (imposed in their paper) is used in the last equality.

Note that a slightly weaker version of Condition \ref{condi_2} can
be stated by replacing $\eta$ with $T$ and the integral with a summation.
Then, analogous to Theorem \ref{thm:model2}, one can readily show
that $F_{U_{1}|T,Z}=F_{U_{0}|T,Z}$ implies such a condition.\end{remark}

\section{Systematic Calculation of Bounds\label{sec:Systematic-Calculation-of}}

\noindent In Theorem \ref{thm:main}, $\gamma$ is required to satisfy
a set of linear inequality constraints, i.e., \eqref{eq:counter_k}
(respectively, \eqref{eq:counter_k2}), and each feasible $\gamma$
establishes an upper bound (respectively, lower bound) on $F_{Y_{1}|D,X}(\cdot|0,x)$.
Therefore, it is intuitive to employ optimization methods to calculate
these bounds, as detailed in Corollary \ref{corol:main}. For simplicity,
our subsequent discussion will focus solely on the upper bound, omitting
covariates $X$ for brevity.

\subsection{Semi-Infinite Programming}

To simplify notation, let $\boldsymbol{p}(y,d)\equiv(p(y,d|z_{1}),...,p(y,d|z_{L}))'$
where $p(y,d|z_{\ell})\equiv P[Y\le y,D=d|Z=z_{\ell}]$ and $p(y|d)\equiv P[Y\le y|D=d]$.
Also, let $\boldsymbol{1}\equiv(1,...,1)'$ and $\boldsymbol{p}\equiv(p(z_{1}),...,p(z_{L}))'$
with $p(z)\equiv P[D=1|Z=z]$ so that 
\begin{align*}
\Gamma & =\{\gamma:\gamma'[\begin{array}{cc}
\boldsymbol{1} & \boldsymbol{p}\end{array}]=[\begin{array}{cc}
0 & 1\end{array}]\}\subset\mathbb{R}^{L}.
\end{align*}
Consider the following linear semi-infinite programming problem for
the upper bound on $P[Y_{0}\leq\bar{y}|D=1]$:
\begin{align}
UB(\bar{y}) & =\min_{\gamma\in\Gamma_{p}}-\boldsymbol{p}(\bar{y},0)'\gamma\label{LP1}\\
s.t. & \quad\boldsymbol{p}(y,1)'\gamma\ge p(y|1),\quad\forall y\in\mathcal{Y}\label{LP2}
\end{align}
Note that the existence of $\gamma$ satisfying condition \eqref{eq:counter_k}
guarantees that the feasible set is non-empty. Also note that this
condition is allowed to satisfy only almost everywhere (a.e.), which
we suppress for simplicity. This program is infeasible to solve in
practice as there are infinitely many constraints. We propose two
approaches to approximate it with a linear program (LP).

\subsection{Linear Program with Randomized Constraints\label{subsec:Linear-Program-with}}

One approach to the semi-infinite program \eqref{LP1}--\eqref{LP2}
is to approximate \eqref{LP2} by 
\[
\boldsymbol{p}(\tilde{Y}_{m},1)'\gamma\ge p(\tilde{Y}_{m}|1),a.s.\ \text{for }m=1,\cdots,s_{n},
\]
where $\{\tilde{Y}_{m}:m=1,\cdots,s_{n}\}$ is an i.i.d. simulated
sample as is done in the literature (e.g., \citet{calafiore2005uncertain}).
An obvious candidate of this sample would be $\{Y_{i}\}_{i=1}^{n}$
with $s_{n}=n$. Consider a sampled LP of the following:

\begin{align}
\overline{UB}_{n}(\bar{y}) & =\min_{\gamma\in\Gamma}-\boldsymbol{p}(\bar{y},0)'\gamma\label{sLP1}\\
s.t. & \quad\boldsymbol{p}(Y_{i},1)'\gamma\ge p(Y_{i}|1).\quad\forall i=1,...,n\label{sLP2}
\end{align}
In Section \ref{sec:Bounding-Violation-Probability} of the Appendix,
we show that the probability of violating the original constraints
\eqref{LP2} by using \eqref{sLP2} can be bounded by $O(1/n)$.

\subsection{Dual Program and Sieve Approximation\label{subsec:Dual-Program-and}}

Another approach to the semi-infinite program \eqref{LP1}--\eqref{LP2}
is to invoke its dual and approximate the Lagrangian measure using
sieve. With the constraint $p(\cdot|1)-\boldsymbol{p}(\cdot,1)'\gamma\le0$,
the Lagrangian for \eqref{LP1}--\eqref{LP2} can be written as 
\begin{align*}
\mathcal{L}(\bar{y},\gamma,\Lambda,\lambda) & =-\boldsymbol{p}(\bar{y},0)'\gamma+\int_{\mathcal{Y}}[p(y|1)-\boldsymbol{p}(y,1)'\gamma]d\Lambda(y)+\lambda'([\begin{array}{cc}
\boldsymbol{1} & \boldsymbol{p}\end{array}]'\gamma-[\begin{array}{cc}
0 & 1\end{array}]')\\
 & =\int_{\mathcal{Y}}p(y|1)d\Lambda(y)-[\begin{array}{cc}
0 & 1\end{array}]\lambda+(\lambda'[\begin{array}{cc}
\boldsymbol{1} & \boldsymbol{p}\end{array}]'-\int_{\mathcal{Y}}\boldsymbol{p}(y,1)'d\Lambda(y)-\boldsymbol{p}(\bar{y},0)')\gamma,
\end{align*}
where $\Lambda$ is a non-negative (not necessarily probability) measure
(i.e., $\Lambda\succeq0$) that assigns weights to binding constraints.
Moreover, let 
\begin{align*}
UB(\bar{y}) & =\min_{\gamma\in\mathbb{R}^{L}}\sup_{\Lambda\succeq0,\lambda\in\mathbb{R}^{2}}\mathcal{L}(\bar{y},\gamma,\Lambda,\lambda).
\end{align*}
 Then, we have the following dual problem:

\begin{lemma}\label{lem:dual}The dual problem of the primal problem
of \eqref{LP1}--\eqref{LP2} is given by
\begin{align}
\widetilde{UB}(\bar{y}) & =\sup_{\Lambda\succeq0,\lambda\in\mathbb{R}^{2}}\int_{\mathcal{Y}}p(y|1)d\Lambda(y)-[\begin{array}{cc}
0 & 1\end{array}]\lambda\label{LP1_dual}\\
s.t. & \qquad[\begin{array}{cc}
\boldsymbol{1} & \boldsymbol{p}\end{array}]\lambda-\int_{\mathcal{Y}}\boldsymbol{p}(y,1)d\Lambda(y)-\boldsymbol{p}(\bar{y},0)=\boldsymbol{0}.\label{LP2_dual}
\end{align}
\end{lemma}

Note that \eqref{LP2_dual} has a finite number of constraints (i.e.,
$L$ constraints). It is trivial to show weak duality, $\widetilde{UB}(\bar{y})\le UB(\bar{y})$.\footnote{This is because, by \eqref{LP1}--\eqref{LP2} and \eqref{LP1_dual}--\eqref{LP2_dual},
we have
\begin{align*}
-\boldsymbol{p}(\bar{y},0)'\gamma & =\left\{ \int_{\mathcal{Y}}\boldsymbol{p}(y,1)d\Lambda(y)-[\begin{array}{cc}
\boldsymbol{1} & \boldsymbol{p}\end{array}]\lambda\right\} '\gamma=\int_{\mathcal{Y}}\boldsymbol{p}(y,1)'\gamma d\Lambda(y)-\lambda'[\begin{array}{cc}
\boldsymbol{1} & \boldsymbol{p}\end{array}]'\gamma\ge\int_{\mathcal{Y}}p(y|1)d\Lambda(y)-\lambda'[\begin{array}{cc}
0 & 1\end{array}]'.
\end{align*}
} Strong duality also holds because of the structure of the problem
(i.e., linearity in $\gamma$, continuity of $\boldsymbol{p}(\cdot,d)$
and $p(\cdot|d)$). We establish this in the following theorem.

\begin{myas}{C}\label{as:C}$\mathcal{Y}$ is compact.\end{myas}

\begin{theorem}\label{thm:strong_duality}Suppose Assumption \ref{as:C}
holds and there is $\gamma^{*}\in\{y:\boldsymbol{p}(y,1)'\gamma\ge p(y|1)\}$
such that $\boldsymbol{p}(y,1)'\gamma^{*}>p(y|1)$. Then, if the primal
solution $UB(\bar{y})$ is finite, then $\widetilde{UB}(\bar{y})=UB(\bar{y})$.\end{theorem}

Note that $\Lambda(y)$ is smooth as the feasible set of the primal
problem is smooth due to the smoothness of $p(y|d)$ and $\boldsymbol{p}(y,d)$,
which are CDFs. This motivates us to use sieve approximation for $\Lambda(y)$
to turn the dual into a linear programming problem. The smoothness
class for $\Lambda(y)$ will be determined by the smoothness class
of CDFs. Let $\mathcal{Y}$ is normalized to be $[0,1]$ and $\lambda(y)\equiv d\Lambda(y)/dy$
that satisfies $\int_{\mathcal{Y}}\lambda(y)dy=1$ and $\lambda(y)\ge0$
for all $y\in\mathcal{Y}$. Consider the following sieve approximation:
\begin{align*}
\lambda(y) & \approx\sum_{j=1}^{J}\theta_{j}b_{j}(y),
\end{align*}
where $b_{j}(y)\equiv b_{j,J}(y)\equiv\left(\begin{array}{c}
J\\
j
\end{array}\right)y^{j}(1-y)^{J-j}$ is a Bernstein basis function. Then, the LP can be written as
\begin{align*}
\widetilde{UB}_{J}(\bar{y}) & =\max_{\theta\in\mathbb{R}_{+}^{J},\lambda\in\mathbb{R}^{2}}\sum_{j=1}^{J}\theta_{j}b_{1,j}-[\begin{array}{cc}
0 & 1\end{array}]\lambda\\
s.t. & \qquad[\begin{array}{cc}
\boldsymbol{1} & \boldsymbol{p}\end{array}]\lambda-\sum_{j=1}^{J}\theta_{j}\boldsymbol{b}_{1,j}-\boldsymbol{p}(\bar{y},0)=\boldsymbol{0},\\
 & \qquad\frac{1}{J+1}\sum_{j=1}^{J}\theta_{j}=1,
\end{align*}
or equivalently,
\begin{align}
\widetilde{UB}_{J}(\bar{y}) & =\max_{\theta\in\mathbb{R}_{+}^{J},\lambda\in\mathbb{R}^{2}}\theta'b_{1}-[\begin{array}{cc}
0 & 1\end{array}]\lambda\label{LP1_dual_approx}\\
s.t. & \qquad[\begin{array}{cc}
\boldsymbol{1} & \boldsymbol{p}\end{array}]\lambda-B_{1}\theta-\boldsymbol{p}(\bar{y},0)=\boldsymbol{0},\label{LP2_dual_approx}\\
 & \qquad\boldsymbol{1}'\theta-J-1=0,\label{LP3_dual_approx}
\end{align}
where $\theta\equiv(\theta_{1},...,\theta_{J})'$, $b_{d}\equiv(b_{d,1},...,b_{d,J})'$
with $b_{d,j}\equiv\int_{\mathcal{Y}}b_{j}(y)p(y|d)dy$, $\boldsymbol{b}_{d,j}\equiv(b_{d,j,1},...,b_{d,j,L})'$
with $b_{d,j,\ell}\equiv\int_{\mathcal{Y}}b_{j}(y)p(y,d|z_{\ell})dy$,
and $B_{d}\equiv[\begin{array}{ccc}
\boldsymbol{b}_{d,1} & \cdots & \boldsymbol{b}_{d,J}\end{array}]$ is an $L\times J$ matrix, and by using $\int_{\mathcal{Y}}b_{j}(y)dy=\frac{1}{J+1}$
for all $j$. Note that the nonnegativity restriction on $\theta$
is imposed to reflect that $\lambda$ is a nonnegative measure. Using
Bernstein polynomials to approximate infinite-dimensional decision
variables is also used in \citet{han2020sharp}.

\begin{remark}[Local Approximation]\label{rem:local_approx}The LP
\eqref{LP1_dual_approx}--\eqref{LP2_dual_approx} may be more stable
than the LP \eqref{sLP1}--\eqref{sLP2}. In terms of dual, the latter
approach is equivalent to having $\sum_{i=1}^{n}p(Y_{i}|1)\lambda_{i}$
as an approximation for $\int_{\mathcal{Y}}p(y|1)\lambda(y)dy$. This
can be viewed as a crude local approximation that involves a uniform
kernel.\end{remark}

\section{Numerical Studies\label{sec:Numerical-Studies}}

\begin{figure}
\centering{}\hspace{-0.7cm}\includegraphics[scale=0.2]{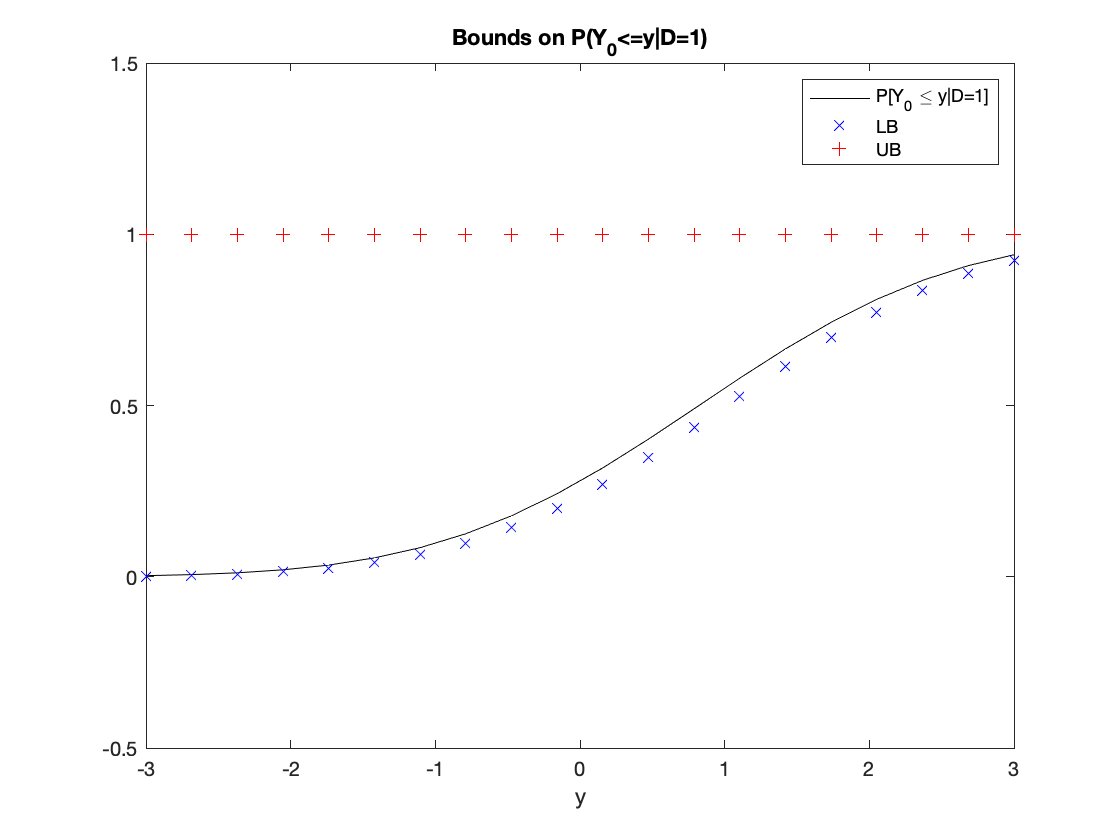}\caption{Bounds on $\Pr[Y_{0}\le y|D=1]$ When $L=2$}
\label{fig:bds_L2}
\end{figure}

\begin{figure}
\centering{}\hspace{-0.7cm}\includegraphics[scale=0.2]{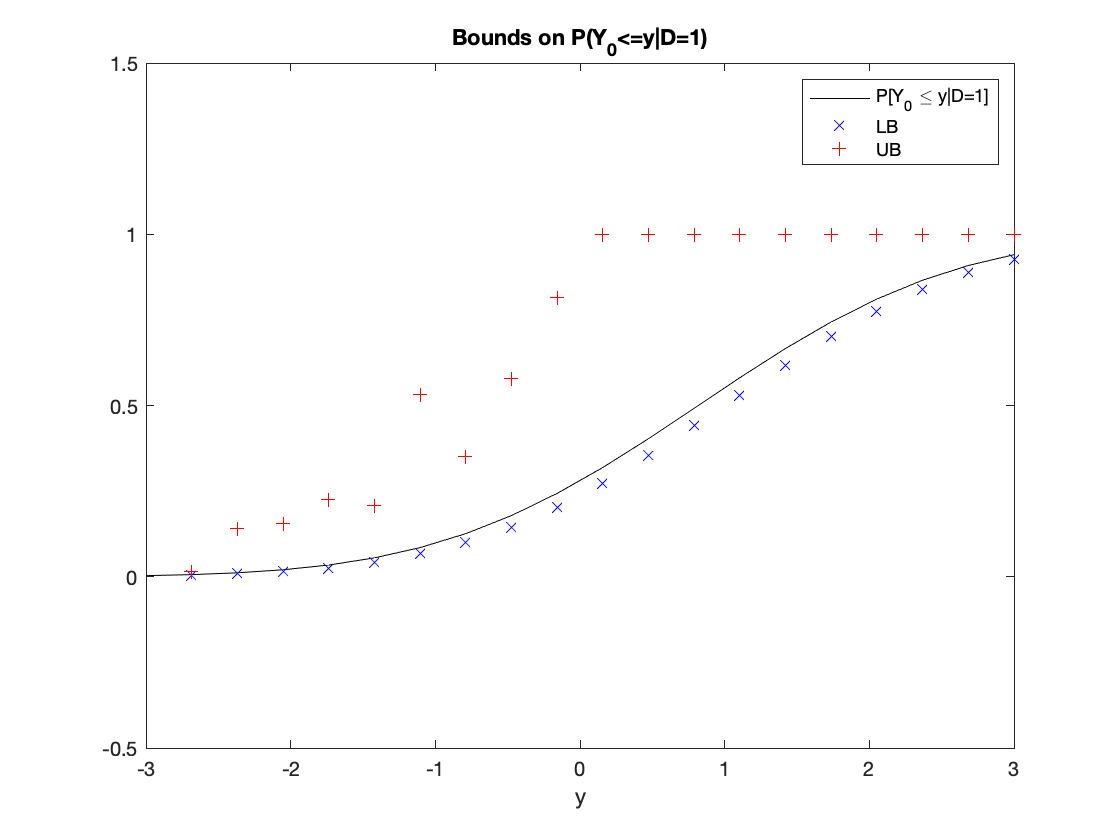}\caption{Bounds on $\Pr[Y_{0}\le y|D=1]$ When $L=5$}
\label{fig:bds_L5}
\end{figure}

\begin{figure}
\centering{}\hspace{-0.7cm}\includegraphics[scale=0.2]{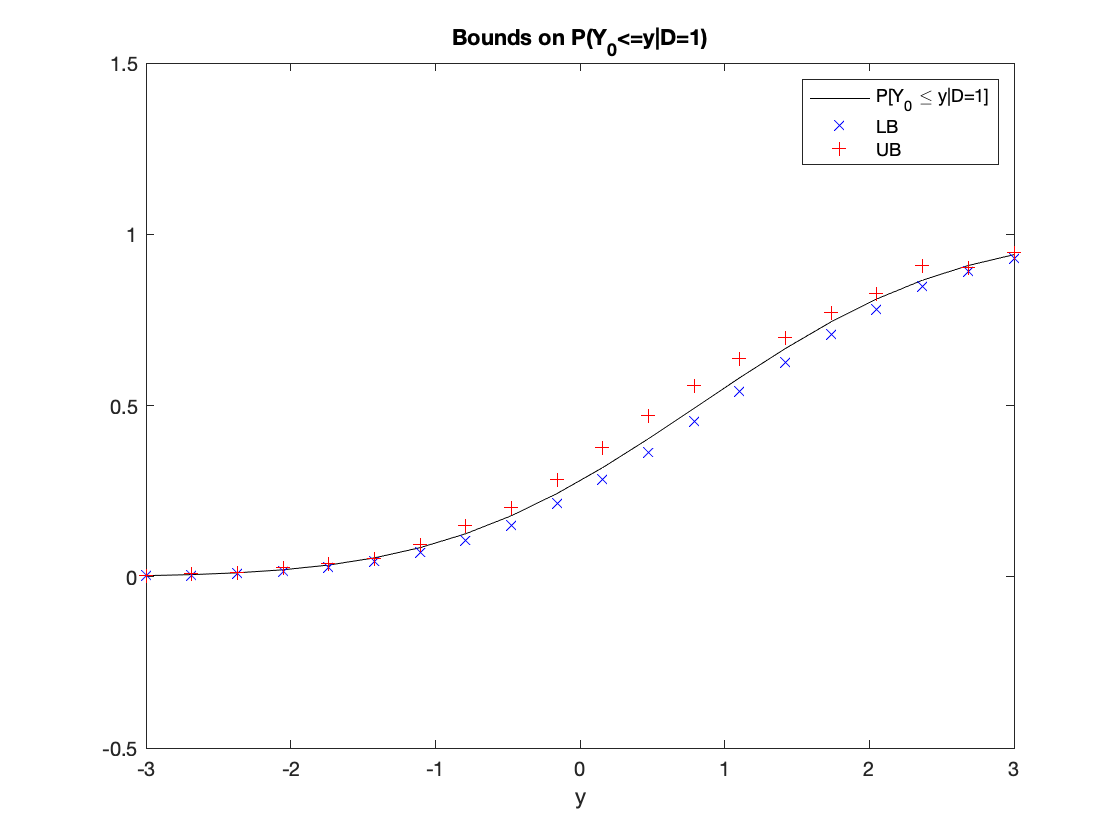}\caption{Bounds on $\Pr[Y_{0}\le y|D=1]$ When $L=6$}
\label{fig:bds_L6}
\end{figure}

To illustrate the importance of multiple IVs and the informativeness
of resulting bounds, we conduct numerical exercises. We generate the
data so that they are consistent with Model 1 and hence satisfy Condition
\ref{condi_1_suff}. The variables $(Y,D,Z)$ are generated in the
following fashion:
\begin{itemize}
\item $Y_{d}=q(d,U_{d})=1-d+(d+1)U_{d}$ for $\mathcal{Y}=\mathbb{R}$,
that is, $Y_{1}=2U_{1}$ and $Y_{0}=1+U_{0}$
\item $(U,\eta)\sim BVN((0,0)',\Sigma)$
\item $V\sim N(0,\sigma_{V}^{2})$ and $\xi_{1}\sim N(0,\sigma_{\xi}^{2})$
\item $\xi_{0}=\xi_{1}+V$
\item $U_{d}=U+\xi_{d}$
\item $Z\sim Bin(L-1,p)/(L-1)\in[0,1]$ with $L\in\{2,3,4,5,6,7,8\}$
\item $D=1\{\pi_{0}+\pi_{1}Z\ge\eta\}$
\item $Y=DY_{1}+(1-D)Y_{0}$
\end{itemize}
Here, $Z$ is normalized so that the endpoints of the support are
invariant regardless of the value of $L$. This is intended to understand
the role of the number of values $Z$ takes while fixing the role
of instrument strength. Figures \ref{fig:bds_L2}--\ref{fig:bds_L6}
presents the bounds on $\Pr[Y_{0}\le y|D=1]$ while varying $L$.
The bounds are calculated using the approach proposed in Section \ref{subsec:Linear-Program-with}.
We only report $L\in\{2,5,6\}$ for succinctness. In these figures,
the black solid line indicates the true value of $\Pr[Y_{0}\le y|D=1]$
and the red and blue crosses depict the upper and lower bounds. Although
the upper bound is a trivial upper bound for the CDF when $L=2$,
it quickly becomes informative as $L$ increases beyond $5$. To put
this in a context, this corresponds to the number of instrument values
that three binary IVs can easily surpass or a single continuous IV.

\newpage{}

\clearpage{}

\begin{appendix}

\section{Point Identification\label{sec:Point-Identification}}

Point identification of $QTE_{\tau}(d,x)$ and $ATE(d,x)$ can be
achieved as long as the stochastic dominance ordering is preserved
(i.e., Condition \ref{as:condi_1} or \ref{as:condi_0}) and instruments
have sufficient variation in a specific sense. As is clear below,
however, we do \emph{not} require $p(z)\rightarrow1$ or $0$ (i.e.,
instruments with large support). In this sense, our approach to point
identification complements the approach of identification at infinity
(e.g., \citet{heckman1990varieties}). To see this, consider the following
theorem.

\begin{theorem}\label{thm:main_point_ID}Suppose that Assumption
\ref{as:Z} and Condition \ref{as:condi_1} hold. Fix $x\in\mathcal{X}$.
For $\gamma\equiv(\gamma_{1},...,\gamma_{L})$ in $\Gamma(x)$, suppose
\begin{align}
 & P[Y\leq\cdot|D=1,X=x]=\sum_{\ell=1}^{L}\gamma_{\ell}P[Y\leq\cdot,D=1|Z=z_{\ell},X=x].\label{eq:counter_k_point_ID}
\end{align}
Then $F_{Y_{0}|D,X}(\cdot|1,x)$ is identified as
\begin{align}
P[Y_{0}\leq\cdot|D=1,X=x] & =-\sum_{\ell=1}^{L}\gamma_{\ell}P[Y\leq\cdot,D=0|Z=z_{\ell},X=x]\label{eq:point_ID}
\end{align}
\end{theorem}

The key for this point identification result is that there exists
$\gamma$ such that \eqref{eq:counter_k_point_ID} holds, which is
a stronger requirement than the inequality version \eqref{eq:counter_k}.
The equation \eqref{eq:counter_k_point_ID} is more likely to hold
when $L$ is large, that is, when instruments take more values. In
particular, when $L\rightarrow\infty$ (e.g., continuous $Z$), we
may view that $P[Y\leq y|D=1,X=x]$ is approximated as
\begin{align*}
P[Y\leq y|D=1,X=x] & =\lim_{L\rightarrow\infty}\sum_{\ell=1}^{L}\gamma_{\ell,L}p(z_{\ell},x)P[Y\leq y|D=1,Z=z_{\ell},X=x],
\end{align*}
where $\sum_{\ell=1}^{L}\gamma_{\ell,L}=0$. Note that, although this
does \emph{not} demand an infinite support for $Z$, it implicitly
assumes that $Z$ sufficiently influences the distribution of $Y$
conditional on $(D,X)=(1,x)$ in a way that the resulting functions,
$P[Y\leq y|D=1,Z=z_{\ell},X=x]$, generate $P[Y\leq y|D=1,X=x]$.
Importantly, whether this is possible or not can be confirmed from
the data.

Given Theorem \ref{thm:main_point_ID}, we identify $QTE_{\tau}(1,x)=Q_{Y|D,X}(\tau|1,x)-Q_{Y_{0}|D,X}(\tau|1,x)$
where $Q_{Y_{0}|D,X}(\tau|1,x)$ is a solution to $\tau=-\sum_{\ell=1}^{L}\gamma_{\ell}P[Y\leq\cdot,D=0|Z=z_{\ell},X=x]$.
Similarly, under Condition \ref{as:condi_0}, we can identify $F_{Y_{0}|D,X}(\cdot|1,x)$
and thus $QTE_{\tau}(0,x)$. We omit this result for succinctness.

It is worth comparing the point identification result with that in
\citet{chernozhukov2005iv}. The latter point identifies $QTE_{\tau}(x)$
with a binary instrument by assuming rank similarity. The result of
this section suggests that the identification of $QTE_{\tau}(x)$
can alternatively be achieved when Conditions \ref{as:condi_1} and
\ref{as:condi_0} both hold and the IVs satisfy \eqref{eq:counter_k_point_ID}.
To see the connection to rank similarity, note that rank similarity
implies Condition \ref{condi_2} (by Theorem \ref{thm:model2}), but
the latter implies Conditions \ref{as:condi_1} and \ref{as:condi_0}
that identify $QTE_{\tau}(1,x)$ and $QTE_{\tau}(0,x)$, respectively,
and thus $QTE_{\tau}(x)$ jointly. In this way, the two approaches
enjoy different levels of the trade-off between restrictions on the
heterogeneity and exogenous variation.

\section{Conditions for Average Treatment Effects\label{sec:Conditions-for-Average}}

To calculate bounds on $ATE(1,x)$ and $ATE(0,x)$, we introduce
conditions that are weaker that Conditions \ref{as:condi_1} and \ref{as:condi_0}.

\begin{mycondi}{S$_1^{\prime}$}\label{as:condi_mean}For arbitrary
non-negative weight vectors $(w_{1},...,w_{L})$ and $(\tilde{w}_{1},...,\tilde{w}_{L})$
that satisfy $\sum_{\ell=1}^{L}w_{\ell}=\sum_{\ell=1}^{L}\tilde{w}_{\ell}=1$,
if
\begin{align}
\sum_{\ell=1}^{L}w_{\ell}P[Y_{1}\le\cdot|D=1,Z=z_{\ell},X=x] & \le\sum_{\ell=1}^{L}\tilde{w}_{\ell}P[Y_{1}\le\cdot|D=1,Z=z_{\ell},X=x],\label{eq:condi_mean_0}
\end{align}
then
\begin{align}
\sum_{\ell=1}^{L}w_{\ell}E[Y_{0}|D=1,Z=z_{\ell},X=x] & \le\sum_{\ell=1}^{L}\tilde{w}_{\ell}E[Y_{0}|D=1,Z=z_{\ell},X=x].\label{eq:condi_mean_1}
\end{align}

\end{mycondi}Condition \ref{as:condi_mean} can be used to bound
the $ATE(1,x)$. An analogous condition can be imposed to bound $ATE(0,x)$.

\begin{mycondi}{S$_0^{\prime}$}\label{as:condi_mean-1}For arbitrary
non-negative weight vectors $(w_{1},...,w_{L})$ and $(\tilde{w}_{1},...,\tilde{w}_{L})$
that satisfy $\sum_{\ell=1}^{L}w_{\ell}=\sum_{\ell=1}^{L}\tilde{w}_{\ell}=1$,
if
\begin{align}
\sum_{\ell=1}^{L}w_{\ell}P[Y_{0}\le\cdot|D=1,Z=z_{\ell},X=x] & \le\sum_{\ell=1}^{L}\tilde{w}_{\ell}P[Y_{0}\le\cdot|D=1,Z=z_{\ell},X=x],\label{eq:condi_mean_0-1}
\end{align}
then
\begin{align}
\sum_{\ell=1}^{L}w_{\ell}E[Y_{1}|D=1,Z=z_{\ell},X=x] & \le\sum_{\ell=1}^{L}\tilde{w}_{\ell}E[Y_{1}|D=1,Z=z_{\ell},X=x].\label{eq:condi_mean_1-1}
\end{align}

\end{mycondi}

\section{Other Structural Models as Sufficient Conditions\label{sec:Other-Structural-Models}}

We present two more structural models that are not nested to Model
1 in the text. Model1(i) are maintained in these models, that is,
$Y=q(D,X,U_{D})$ where $q(d,x,\cdot)$ is continuous and monotone
increasing and $D=h(Z,X,\eta)$.

\bigskip{}

\noindent \textbf{Model 2.} (ii) $U_{0}\stackrel{d}{=}\phi(U_{1},V)$
conditional on $(\eta,X)$ where $V\perp(U_{1},\eta)|X$ and $\phi(\cdot,v)$
is strictly increasing for all $v$.\bigskip{}

Model 2(ii) defines that $U_{0}$ is ``noisier'' than $U_{1}$.
Therefore, Model 2 is weaker than the model in \citet{chernozhukov2005iv}.
Model 2 and Model 1 are not nested because, in $U_{0}=U_{1}+V$ of
Model 1, $V$ is not independent of $U_{1}$. We show below that Model
2 implies Condition \ref{as:condi_1}. Interestingly, Model 2(ii)
with $U_{0}=\phi(U_{1},V)$ (instead of ``$\stackrel{d}{=}$'')
is a generalization of the definition that $U_{0}$ is ``noisier''
than $U_{1}$ if $U_{0}=U_{1}+V$ with $U_{1}\perp V$ in \citet[p. 1880]{pomatto2020stochastic}.

\bigskip{}

\noindent \textbf{Model 3.} (ii)  $U_{0}\stackrel{d}{=}\max\{\phi(U_{1}),V\}$
conditional on $(\eta,X)$ where $V\perp(U_{1},\eta)|X$ and $\phi(\cdot)$
is strictly increasing.\bigskip{}

We show below that Model 3 implies rank linearity. Model 3 can alternatively
be defined as follows: $Y_{0}\stackrel{d}{=}\max\{\phi(Y_{1}),V\}$
conditional on $(\eta,X)$ where $V\perp(Y_{1},\eta)|X$ and $\phi(\cdot)$
is strictly increasing. Then, this model also implies rank linearity
with $\psi(\cdot)=\phi^{-1}(\cdot)$ because
\begin{align*}
\Pr[Y_{0}\le y|\eta,X] & =\Pr[\phi(Y_{1})\le y,V\le y|\eta,X]=\Pr[Y_{1}\le\phi^{-1}(y)|\eta,X]\Pr[V\le y|X].
\end{align*}
This model provides another interpretation of an insurance policy
($D=1$) as $Y_{1}=\max\{Y_{0},V\}$ guarantees at least $Y_{0}$.
Models 2 and 3 are not nested.

\begin{lemma}\label{lem:models23}(i) Model 2 implies Condition \ref{condi_1_suff};
(ii) Model 3 implies rank linearity.

\end{lemma}

The proof of this lemma is contained in Section \ref{sec:Proofs}.

\section{Bounding Violation Probability in Linear Program with Randomized
Constraints\label{sec:Bounding-Violation-Probability}}

Let $h(\gamma,y)\equiv p(y|1)-\boldsymbol{p}(y,1)'\gamma$. Following
\citet{calafiore2005uncertain}, define a violation probability and
a robustly feasible solution.

\begin{definition}[Violation probability] Let $\gamma\in\Gamma$
be a candidate solution for \eqref{LP1}--\eqref{sLP2}. The probability
of violation of $\gamma$ is defined as 
\begin{align*}
V(\gamma) & =\mathbb{P}\{Y\in\mathcal{Y}:h(\gamma,Y)>0\},
\end{align*}
where $\{Y\in\mathcal{Y}:h(\gamma,Y)>0\}$ is assumed to be measurable.
\end{definition}

Note that $V(\gamma^{*})=0$ where $\gamma^{*}$ is the solution to
\eqref{LP1}--\eqref{sLP2}.

\begin{definition}[$\epsilon$-level solution] For $\epsilon\in[0,1]$,
$\gamma\in\Gamma$ is an $\epsilon$-level robustly feasible solution
if $V(\gamma)\le\epsilon$. \end{definition}

Then, we can show that the violation probability at the solution,
denoted as $\bar{\gamma}_{n}$, to \eqref{sLP1}--\eqref{sLP2} is
on average bounded by $1/n$.

\begin{proposition}\label{prop:violation} Let $\bar{\gamma}_{n}$
be the solution to \eqref{sLP1}--\eqref{sLP2}. Then, 
\begin{align*}
\mathbb{E}_{P^{n}}[V(\bar{\gamma}_{n})] & \le\frac{1}{n+1},
\end{align*}
where $P^{n}$ is the probability measure in the space $\mathcal{Y}^{n}$
of the multi-sample extraction $Y_{1},...,Y_{n}$. \end{proposition}

\begin{corollary} Fix $\epsilon\in[0,1]$ and $\beta\in[0,1]$ and
let 
\begin{align*}
n & \ge\frac{1}{\epsilon\beta}-1.
\end{align*}
Then, with probability no smaller than $1-\beta$, the sampled LP
\eqref{sLP1}--\eqref{sLP2} returns an optimal solution $\hat{\gamma}_{n}$
which is $\epsilon$-level robustly feasible. \end{corollary}

The above results implicitly assume a particular rule of tie-breaking
when there are multiple solutions in the sampled LP (see Theorem 3
in \citet{calafiore2005uncertain}). There is also discussions on
no solution in the paper.

\section{Proofs\label{sec:Proofs}}

\subsection{Proof of Lemma \ref{lem:suff}}

Let $p(z,x)\equiv P[D=1|Z=z,X=x]$ and let $H(z,x)\equiv\{\eta:h(z,x,\eta)=1\}$
be a level set. Then,
\begin{align*}
\sum_{\ell}w_{\ell}P[Y_{1}\le y|D=1,Z=z_{\ell},X=x] & =\sum_{\ell}w_{\ell}P[Y_{1}\le y|\eta\in H(z_{\ell},x),X=x]\\
 & =\int\frac{\sum_{\ell}w_{\ell}1[t\in H(z_{\ell},x)]}{p(z_{\ell},x)}P[Y_{1}\le y|\eta=t,X=x]dt.
\end{align*}
Take $w(t,x)=\frac{\sum_{\ell}w_{\ell}1[t\in H(z_{\ell},x)]}{p(z_{\ell},x)}$.
Then, $w(t,x)$ satisfies
\begin{align*}
\int\frac{\sum_{\ell}w_{\ell}1[t\in H(z_{\ell},x)]}{p(z_{\ell},x)}dt & =1.
\end{align*}
The same argument applies to $\tilde{w}$ and $\tilde{w}(t,x)$, and
also for the distribution of $Y_{0}$. $\square$

\subsection{Proof of Theorem \ref{thm:main}}

We suppress $X$ for simplicity and prove the upper bound; the lower
bound can be analogously derived. Without loss of generality, for
some $\ell^{*}\leq L$, let $\gamma_{\ell}\leq0$ for $\ell\leq\ell^{*}$
and $\gamma_{\ell}>0$ for $\ell>\ell^{*}$. Let $q(z_{\ell})\equiv P[Z=z_{\ell}|D=1]$.
Then, \eqref{eq:counter_k} can be rewritten as
\begin{multline*}
\sum_{\ell=1}^{L}q(z_{\ell})\times P[Y\leq y|D=1,Z=z_{\ell}]-\sum_{\ell=1}^{\ell^{*}}\gamma_{\ell}p(z_{\ell})\times P[Y\leq y|D=1,Z=z_{\ell}]\\
\le\sum_{\ell=\ell^{*}+1}^{L}\gamma_{\ell}p(z_{\ell})\times P[Y\leq y|D=1,Z=z_{\ell}].
\end{multline*}
Let $a\equiv1-\sum_{\ell=1}^{\ell^{*}}\gamma_{\ell}p(z_{\ell})$.
By definition and that $\sum_{\ell=1}^{L}\gamma_{\ell}p(z_{\ell})=1$,
we have $a=\sum_{\ell=\ell^{*}+1}^{L}\gamma_{\ell}p(z_{\ell})$. Therefore,
we have 
\begin{multline*}
\sum_{\ell=1}^{\ell^{*}}\frac{q(z_{\ell})-\gamma_{\ell}p(z_{\ell})}{a}\times P[Y_{1}\leq y|D=1,Z=z_{\ell}]+\sum_{\ell=\ell^{*}+1}^{L}\frac{q(z_{\ell})}{a}\times P[Y_{1}\leq y|D=1,Z=z_{\ell}]\\
\le\sum_{\ell=\ell^{*}+1}^{L}\frac{\gamma_{\ell}p(z_{\ell})}{a}\times P[Y_{1}\leq y|D=1,Z=z_{\ell}],
\end{multline*}
where $\sum_{\ell=1}^{\ell^{*}}\frac{q(z_{\ell})-\gamma_{\ell}p(z_{\ell})}{a}+\sum_{\ell=\ell^{*}+1}^{L}\frac{q(z_{\ell})}{a}=1$
and $\sum_{\ell=\ell^{*}+1}^{L}\frac{\gamma_{\ell}p(z_{\ell})}{a}=1$.
Therefore, by Condition \ref{as:condi_1}, we have 
\begin{multline*}
\sum_{\ell=1}^{k}\frac{q(z_{\ell})-\gamma_{\ell}p(z_{\ell})}{a}\times P[Y_{0}\leq y|D=1,Z=z_{\ell}]+\sum_{\ell=\ell^{*}+1}^{L}\frac{q(z_{\ell})}{a}\times P[Y_{0}\leq y|D=1,Z=z_{\ell}]\\
\le\sum_{\ell=\ell^{*}+1}^{L}\frac{\gamma_{\ell}p(z_{\ell})}{a}\times P[Y_{0}\leq y|D=1,Z=z_{\ell}].
\end{multline*}
Equivalently, we have
\begin{eqnarray*}
P[Y_{0}\leq y|D=1] & \leq & \sum_{\ell=1}^{L}\gamma_{\ell}\times P[Y_{0}\leq y,D=1|Z=z_{\ell}]\\
 & = & \sum_{\ell=1}^{L}\gamma_{\ell}\times\left\{ P[Y_{0}\leq y|Z=z_{\ell}]-P[Y_{0}\leq y,D=0|Z=z_{\ell}]\right\} \\
 & = & \sum_{\ell=1}^{L}\gamma_{\ell}P[Y_{0}\leq y|Z=z_{\ell}]-\sum_{\ell=1}^{L}\gamma_{\ell}\times P[Y_{0}\leq y,D=0|Z=z_{\ell}]\\
 & = & P[Y_{0}\leq y]\times\sum_{\ell=1}^{L}\gamma_{\ell}-\sum_{\ell=1}^{L}\gamma_{\ell}\times P[Y\leq y,D=0|Z=z_{\ell}]\\
 & = & -\sum_{\ell=1}^{L}\gamma_{\ell}\times P[Y\leq y,D=0|Z=z_{\ell}],
\end{eqnarray*}
where the last equality is by $\sum_{\ell=1}^{L}\gamma_{\ell}=0$.
$\square$

\subsection{Proof of Lemma \ref{lem:refutable}}

Note that $P[Y_{d}\le y|D_{z_{1}}=1]=P[Y_{d}\le y|\text{AT}]$ and,
for $\ell=\{2,...,L\}$,
\begin{align*}
P[Y_{d}\le y|D_{z_{\ell}}=1] & =\frac{1}{p(z_{\ell})}P\left[Y_{d}\le y,\{\text{AT}\}\cup\bigcup_{\ell'=2}^{\ell}\{(z_{\ell'-1},z_{\ell'})\text{-C}\}\right]
\end{align*}
by Assumption \ref{as:D*} and $p(z_{\ell})=P[D_{z_{\ell}}=1]$. Then,
in \eqref{eq:condi1_2},

\begin{align*}
 & \sum_{\ell=1}^{L}w_{\ell}P[Y_{d}\le y|D_{z_{\ell}}=1]\\
 & =w_{1}P[Y_{d}\le y|\text{AT}]+\sum_{\ell=2}^{L}\frac{w_{\ell}}{p(z_{\ell})}\left(p_{1}P[Y_{d}\le y|\text{AT}]+\sum_{\ell'=2}^{\ell}p_{\ell'}P[Y_{d}\le y|(z_{\ell'-1},z_{\ell'})\text{-C}]\right)
\end{align*}
and similarly for the right-hand side of \eqref{eq:condi1_2}. This
proves (i). To remove the distributions for AT in the expressions,
we set
\begin{align}
w_{1}+p_{1}\sum_{\ell=2}^{L}\frac{w_{\ell}}{p(z_{\ell})} & =\tilde{w}_{1}+p_{1}\sum_{\ell=2}^{L}\frac{\tilde{w}_{\ell}}{p(z_{\ell})}.\label{eq:dof-1}
\end{align}
Then, note that when $L\ge2$, $w\neq\tilde{w}$ even if $w$ and
$\tilde{w}$ satisfy \eqref{eq:dof-1}. Therefore, the resulting \eqref{eq:condi1_2}
is the dominance between the two distinct weight sums of $P[Y_{d}\le y|(z_{\ell'-1},z_{\ell'})\text{-C}]$'s:
\begin{align*}
\sum_{\ell=2}^{L}\frac{w_{\ell}}{\sum_{\ell'=1}^{\ell}p_{\ell'}}\sum_{\ell'=2}^{\ell}p_{\ell'}P[Y_{d}\le y|(z_{\ell'-1},z_{\ell'})\text{-C}] & \le\sum_{\ell=2}^{L}\frac{\tilde{w}_{\ell}}{\sum_{\ell'=1}^{\ell}p_{\ell'}}\sum_{\ell'=2}^{\ell}p_{\ell'}P[Y_{d}\le y|(z_{\ell'-1},z_{\ell'})\text{-C}],
\end{align*}
which can be simplified as \eqref{eq:refutable} in (ii). $\square$

\subsection{Proof of Theorem \ref{thm:model1}}

We suppress $X$ for simplicity. For an arbitrary r.v. $A$, let $F_{A}^{w}(\cdot)\equiv\int w(t)F_{A|\eta}(\cdot|t)dt$,
which itself is a CDF. By \eqref{eq:model1-1} in Model 1(i), $F_{Y_{d}}^{w}\leq F_{Y_{d}}^{\tilde{w}}$
if and only if $F_{U_{d}}^{w}\leq F_{U_{d}}^{\tilde{w}}$ . So it
suffices to show that, if $F_{U_{1}}^{w}\leq F_{U_{1}}^{\tilde{w}}$,
then $F_{U_{0}}^{w}\leq F_{U_{0}}^{\tilde{w}}$.

Let $G(\cdot)$ be an arbitrary monotone increasing function and $g(\cdot)\equiv G'(\cdot)$.
Note that 
\begin{align*}
 & \int GdF_{U_{0}}^{w}-\int GdF_{U_{0}}^{\tilde{w}}=\int[\int\tilde{w}(t)F_{U_{0}|\eta}(u|t)dt-\int w(t)F_{U_{0}|\eta}(u|t)dt]g(u)du\\
 & =\int[\int\tilde{w}(t)\int F_{U|\eta}(u-s|t)f_{\xi_{0}}(s)dsdt-\int w(t)\int F_{U|\eta}(u-s|t)f_{\xi_{0}}(s)dsdt]g(u)du\\
 & =\int\int\int[\tilde{w}(t)-w(t)]F_{U|\eta}(u|t)f_{\xi_{0}}(s)g(u+s)dudsdt,
\end{align*}
where the first eq. is due to the integration by part, the second
eq. is by $F_{U_{d}|\eta}(u|t)=\int F_{U|\eta}(u-s|t)f_{\xi_{0}|\eta}(s|t)ds=\int F_{U|\eta}(u-s|t)f_{\xi_{0}}(s)ds$
under Model 1(ii), and the last eq. is by change of variables. By
Model 1(iii), $f_{\xi_{0}}(s)=\int f_{\xi_{1}}(s-v)f_{V}(v)dv=\int f_{\xi_{1}}(v)f_{V}(s-v)dv$
where $f_{A}(\cdot)$ is the PDF of an arbitrary r.v. $A$. Therefore,
\begin{align*}
 & \int GdF_{U_{0}}^{w}-\int GdF_{U_{0}}^{\tilde{w}}\\
 & =\int\int\int[\tilde{w}(t)-w(t)]F_{U|\eta}(u|t)\int f_{\xi_{1}}(v)f_{V}(s-v)g(u+s)dvdudsdt\\
 & =\int\int[\tilde{w}(t)-w(t)]F_{U|\eta}(u|t)\int f_{\xi_{1}}(v)[\int f_{V}(s)g(u+s+v)ds]dvdudt.
\end{align*}
Let $\psi(s)\equiv\int f_{V}(t)g(t+s)dt$. By definition, $\psi\geq0$
since $g\geq0$. Therefore, 
\begin{align*}
 & \int GdF_{U_{0}}^{w}-\int GdF_{U_{0}}^{\tilde{w}}\\
 & =\int\int[\tilde{w}(t)-w(t)]F_{U|\eta}(u|t)\int f_{\xi_{1}}(v)\psi(u+v)dvdudt\\
 & =\int\int[\tilde{w}(t)-w(t)]\int F_{U|\eta}(u-v|t)f_{\xi_{1}}(v)dv\psi(u)dudt\\
 & =\int\int[\tilde{w}(t)-w(t)]\int F_{U_{1}|\eta}(u|t)\psi(u)dudt\\
 & =\int[\int\tilde{w}(t)F_{U_{1}|\eta}(u|t)dt-\int w(t)F_{U_{1}|\eta}(u|t)dt]\psi(u)du\ge0,
\end{align*}
where the last ineq. is by $F_{U_{1}}^{w}\leq F_{U_{1}}^{\tilde{w}}$.
Because $G(\cdot)$ is arbitrary, then $F_{U_{0}}^{w}$ is first order
stochastic dominant over $F_{U_{0}}^{\tilde{w}}$. $\square$

\subsection{Proof of Theorem \ref{thm:equiv_RL}: Equivalence Between Rank Linearity
and Condition \ref{condi_2}}

The ``if'' part is trivial. We prove ``only if'' part. Suppress
$(Z,X)$ for simplicity. Suppose Condition \ref{condi_2} holds. Let
$\mathcal{Y}_{\infty}\equiv\{y_{k}\in\mathbb{R}:k=1,\cdots,\infty\}$
be a sequence that is dense on $\mathbb{R}$. Denote $\mathcal{Y}_{n}\equiv\{y_{k}\in\mathbb{R}:k=1,\cdots,n\}$.
Because $\mathcal{Y}_{\infty}$ is dense in $\mathbb{R}$ and CDFs
are right-continuous, it suffices to show the existence of $\lambda(\cdot)$
and $\psi(\cdot)$ on $\mathcal{Y}_{\infty}$ such that 
\begin{equation}
F_{Y_{0}|\eta}(\psi(y)|t)=\lambda(y)F_{Y_{1}|\eta}(y|t)
\end{equation}
holds for all $t\in\mathcal{T}$ and $y\in\mathcal{Y}_{\infty}$.

Fix $n\in\mathbb{N}$. Let $G_{1,k}:\mathbb{R}\rightarrow\{0,1\}$
be a simple function defined as $G_{1,k}(\cdot)\equiv1\{y_{k}\le\cdot\}$
for $k=1,\cdots,n$. By the full rank condition \eqref{eq:rank_condi},
for each $1\leq k\leq n$, there exists a function $c_{k}:\mathcal{T}\rightarrow\mathbb{R}$
such that 
\[
G_{1,k}(\cdot)=\int c_{k}(t)F_{Y_{1}|\eta}(\cdot|t)dt.
\]
Define $G_{0,k}:\mathbb{R}\rightarrow[0,1]$ as 
\[
G_{0,k}(\cdot)\equiv\int c_{k}(t)F_{Y_{0}|\eta}(\cdot|t)dt.
\]
Note that $G_{0,k}$ is a proper CDF. Now, for any vectors $\pi\equiv(\pi_{1},\cdots,\pi_{n})$
and $\tilde{\pi}\equiv(\tilde{\pi}_{1},\cdots,\tilde{\pi}_{n})$ such
that $\sum_{k=1}^{n}\pi_{k}=\sum_{k=1}^{n}\tilde{\pi}_{k}=1$, suppose
\[
\sum_{k=1}^{n}\pi_{k}G_{1,k}(\cdot)\leq\sum_{k=1}^{n}\tilde{\pi}_{k}G_{1,k}(\cdot).
\]
It follows that
\[
\int b_{n}(t)F_{Y_{1}|\eta}(\cdot|t)dt\leq\int\tilde{b}_{n}(t)F_{Y_{1}|\eta}(\cdot|t)dt,
\]
where $b_{n}(t)\equiv\sum_{k=1}^{n}\pi_{k}c_{k}(t)$ and $\tilde{b}_{n}(t)\equiv\sum_{k=1}^{n}\tilde{\pi}_{k}c_{k}(t)$.
Let $b_{n}^{+}(t)=\max\{b_{n}(t),0\}$ and $b_{n}^{-}(t)=\min\{b_{n}(t),0\}$
and similarly define $\tilde{b}_{n}^{+}(t)$ and $\tilde{b}_{n}^{-}(t)$.
Then, the above inequality can be written as
\[
\int\{b_{n}^{+}(t)-\tilde{b}_{n}^{-}(t)\}F_{Y_{1}|\eta}(\cdot|t)dt\leq\int\{\tilde{b}_{n}^{+}(t)-b_{n}^{-}(t)\}F_{Y_{1}|\eta}(\cdot|t)dt,
\]
where the resulting weight functions on both sides are non-negative.
Then, by Condition \ref{condi_2}, we have 
\[
\sum_{k=1}^{n}\pi_{k}G_{0,k}(\cdot)\leq\sum_{k=1}^{n}\tilde{\pi}_{k}G_{0,k}(\cdot)
\]
By a similar argument, the converse is also true and thus we have
\[
\sum_{k=1}^{n}\pi_{k}G_{1,k}(\cdot)\leq\sum_{k=1}^{n}\tilde{\pi}_{k}G_{1,k}(\cdot).
\]
if and only if 
\[
\sum_{k=1}^{n}\pi_{k}G_{0,k}(\cdot)\leq\sum_{k=1}^{n}\tilde{\pi}_{k}G_{0,k}(\cdot)
\]
for any non-negative weights $\pi$ and $\tilde{\pi}$. Therefore,
it follows that 
\begin{equation}
\sum_{k=1}^{n}\delta_{k}G_{1,k}(\cdot)\leq0\text{ if and only if }\sum_{k=1}^{n}\delta_{k}G_{0,k}(\cdot)\leq0\label{eq:RL_proof-1}
\end{equation}
for any $n$-dimensional vector $\delta\equiv(\delta_{1},\cdots,\delta_{n})$
that satisfies $\sum_{k=1}^{n}\delta_{k}=0$.

For $d\in\{0,1\}$, define
\begin{align*}
\Delta_{d}^{G} & \equiv\left\{ \delta\in\mathbb{R}^{n}:\sum_{k=1}^{n}\delta_{k}G_{d,k}(y)\leq0\text{ }\forall y\in\mathbb{R};\sum_{k=1}^{n}\delta_{k}=0\right\} .
\end{align*}
Note that $\{\big(G_{1,1}(y),\cdots,G_{1,n}(y)\big):y\in\mathbb{R}\}=\{\big(G_{1,1}(y),\cdots,G_{1,n}(y)\big):y\in\mathcal{Y}_{n}\}$
by definition. Therefore, $\Delta_{1}^{G}$ is a \emph{finite} cone
and its dimension is $n-1$. Define the polar cone of $\Delta_{d}^{G}$
as $\Delta_{d}^{G*}\equiv\{G_{d}\in\mathbb{R}^{n}:G_{d}'\delta\le0,\forall\delta\in\Delta_{d}^{G}\}$.
Note that by definition, $\big(G_{1,1}(y),\cdots,G_{1,n}(y)\big)$
for $y\in\mathcal{Y}_{n}/\{y_{n}\}$ are $n-1$ linearly independent
vectors and therefore generate extreme rays of $\Delta_{1}^{G*}$.
Also note that any element in $\Delta_{0}^{G*}$ is written as $\big(G_{0,1}(y),\cdots,G_{0,n}(y)\big)$
for some $y\in\mathbb{R}$, and so is a vector that generates its
extreme ray. But by \eqref{eq:RL_proof-1}, we have that $\Delta_{1}^{G}=\Delta_{0}^{G}$
and thus $\Delta_{1}^{G*}=\Delta_{0}^{G*}$, and therefore, for each
$y_{k}\in\mathcal{Y}_{n}/\{y_{n}\}$, there exists $y_{k}^{*}\in\mathbb{R}$
and $\lambda_{n}(\cdot)>0$ such that 
\begin{equation}
\big(G_{0,1}(y_{k}^{*}),\cdots,G_{0,n}(y_{k}^{*})\big)=\lambda_{n}(y_{k})\times\big(G_{1,1}(y_{k}),\cdots,G_{1,n}(y_{k})\big).\label{eq:RL_G}
\end{equation}
If there exists multiple values of $y_{k}^{*}$ satisfying \eqref{eq:RL_G},
we define $y_{k}^{*}$ as the infimum of $\{\tilde{y}_{k}^{*}:\big(G_{0,1}(\tilde{y}_{k}^{*}),\cdots,G_{0,n}(\tilde{y}_{k}^{*})\big)=\lambda_{n}(y_{k})\times\big(G_{1,1}(y_{k}),\cdots,G_{1,n}(y_{k})\big)\}$.
Because CDFs are right-continuous function, the infimum should also
satisfy \eqref{eq:RL_G}.

For any $j,k=1,...,n$, if $G_{1,j}(y_{k})=0$ then $G_{0,j}(y_{k}^{*})=0$
by \eqref{eq:RL_G}, which further implies that $G_{0,j}(y^{*})=0$
for all $y^{*}\leq y_{k}^{*}$ because $G_{0,j}(\cdot)$ is monotone
increasing. Let $\{j_{1},\cdots,j_{n}\}$ be a permutation of $\{1,\cdots,n\}$
such that $y_{j_{1}}<y_{j_{2}}<\cdots<y_{j_{n}}$. Note that $G_{1,j_{1}}(y_{j_{1}})$
is the only non-zero component in the set $\{G_{1,k}(y_{j_{1}}):k=1,\cdots,n\}$.
Then, by \eqref{eq:RL_G}, $G_{0,j_{1}}(y_{j_{1}}^{*})\neq0$ and
$G_{0,j_{k}}(y_{j_{1}}^{*})=0$ for $k\geq2$. Similarly, there are
two elements of $\{G_{0,k}(y_{j_{2}}^{*}):k=1,\cdots,n\}$ which are
non-zero, namely, $G_{0,j_{1}}(y_{j_{2}}^{*})$ and $G_{0,j_{2}}(y_{j_{2}}^{*})$.
Therefore, by $G_{0,j_{2}}(y_{j_{1}}^{*})=0$ and $G_{0,j_{2}}(y_{j_{2}}^{*})\neq0$
and the fact that $G_{0,j_{2}}(\cdot)$ is monotone increasing, we
can conclude $y_{j_{1}}^{*}<y_{j_{2}}^{*}$. Continuing this argument,
we can conclude that 
\[
y_{j_{1}}^{*}<y_{j_{2}}^{*}<\cdots<y_{j_{n}}^{*}.
\]
Define a function $\psi_{n}:y_{k}\mapsto y_{k}^{*}$ for $k=1,\cdots,n$.
By the above analysis, $\psi_{n}(\cdot)$ is a monotone increasing
function. Note that the support of $\psi_{n}$ is $\mathcal{Y}_{n}$,
which we extend to $\mathbb{R}$ as follows: for any $y\in\mathbb{R}$,
\[
\psi_{n}(y)=\left\{ \begin{array}{cc}
\max\{\psi_{n}(y_{k}):y_{k}\leq y,\text{ }k=1,\cdots,n\} & \text{ if }y\geq\min\{y_{1},\cdots,y_{n}\}\\
-\infty & \text{otherwise}
\end{array}\right.
\]
Then, $\psi_{n}:\mathbb{R}\rightarrow\mathbb{R}$ is still a monotone
increasing function.

We now consider increasing $n$ to $n+1$. By a similar argument,
there exists a sequence $\{y_{1}^{\dagger},\cdots,y_{n}^{\dagger},y_{n+1}^{\dagger}\}$
and $\lambda_{n+1}(\cdot)>0$ such that for $k=1,\cdots,n+1$, we
have 
\begin{equation}
\big(G_{0,1}(y_{k}^{\dagger}),\cdots,G_{0,n}(y_{k}^{\dagger}),G_{0,n+1}(y_{k}^{\dagger})\big)=\lambda_{n+1}(y_{k})\times\big(G_{1,1}(y_{k}),\cdots,G_{1,n}(y_{k}),G_{1,n+1}(y_{k})\big),\label{eq:RL_G2}
\end{equation}
If there exists multiple values of $y_{k}^{\dagger}$, we define $y_{k}^{\dagger}$
as the infimum of them. Note that, by \eqref{eq:RL_G2} and \eqref{eq:RL_G},
$y_{k}^{\dagger}$ is one of the candidates $\tilde{y}_{k}^{*}$'s
that make $\big(G_{0,1}(\tilde{y}_{k}^{*}),\cdots,G_{0,n}(\tilde{y}_{k}^{*})\big)$
proportional to $\big(G_{1,1}(y_{k}),\cdots,G_{1,n}(y_{k})\big)$
satisfy \eqref{eq:RL_G}. While $y_{k}^{*}$ is the infimum of those
candidates, $y_{k}^{\dagger}$ cannot reach that infimum because it
has to satisfies the additional restriction, $G_{0,n+1}(y_{k}^{\dagger})=\lambda_{n+1}(y_{k})G_{1,n+1}(y_{k})$.
Therefore, we can conclude that $y_{k}^{\dagger}\geq y_{k}^{*}$ for
$k=1,\cdots,n$. Using $\{y_{1},\cdots,y_{n},y_{n+1}\}$ and $\{y_{1}^{\dagger},\cdots,y_{n}^{\dagger},y_{n+1}^{\dagger}\}$,
define $\psi_{n+1}(\cdot)$ analogous to $\psi_{n}(\cdot)$ above.
Then, $\psi_{n+1}(y_{k})=y_{k}^{\dagger}\ge y_{k}^{*}=\psi_{n}(y_{k})$
for $k=1,\cdots,n$. Furthermore, by definition, $\psi_{n+1}(y_{n+1})\ge\psi_{n}(y_{n+1})$
regardless of the rank order of $y_{n+1}$ in $\mathcal{Y}_{n+1}$.
Therefore, for any $y\in\mathbb{R}$,
\[
\psi_{n+1}(y)\geq\psi_{n}(y),
\]
and thus the limit of the sequence of functions $\psi_{n}(\cdot)$
exists as $n\rightarrow\infty$, which we denote as $\psi_{\infty}(\cdot)$.
Recall each $\psi_{n}(\cdot)$ is weakly increasing. It is easy to
prove by contradiction that $\psi_{\infty}(\cdot)$ is strictly increasing.
Fix $y_{k}\in\mathcal{Y}_{\infty}$. For any $n\ge k$, $\big(G_{0,1}(\psi_{\infty}(y_{k})),\cdots,G_{0,n}(\psi_{\infty}(y_{k}))\big)$
is proportional to $\big(G_{1,1}(y_{k}),\cdots,G_{1,n}(y_{k})\big)$
and therefore there exists $\lambda_{\infty}(y_{k})$ such that 
\begin{equation}
\big(G_{0,1}(\psi_{\infty}(y_{k})),\cdots,G_{0,n}(\psi_{\infty}(y_{k}))\big)=\lambda_{\infty}(y_{k})\times\big(G_{1,1}(y_{k}),\cdots,G_{1,n}(y_{k})\big)\label{eq:RL_G3}
\end{equation}
for any $n\in\mathbb{N}$. Moreover, because $\mathcal{Y}_{\infty}$
is dense in $\mathbb{R}$ and $\psi_{\infty}$ and $G_{d,k}$ are
right-continuous functions, the above condition holds for all $y\in\mathbb{R}$.

Note $\{G_{1,k}(\cdot):k=1,\cdots,\infty\}$ is a class of simple
functions. Therefore, any $F_{Y_{1}|\eta}(\cdot|t)$ can be written
as 
\[
F_{Y_{1}|\eta}(\cdot|t)=\lim_{K\rightarrow\infty}\sum_{k=1}^{K}a_{K,k}(t)G_{1,k}(\cdot)
\]
for some triangular array $\{a_{Kk}(t):1\leq k\leq K,K=1,2,\cdots,\infty\}$.
By the definition of $G_{1,k}(\cdot)$, it follows that 
\begin{align}
F_{Y_{1}|\eta}(\cdot|t) & =\lim_{K\rightarrow\infty}\sum_{k=1}^{K}a_{K,k}(t)\int w_{k}(s)F_{Y_{1}|\eta}(\cdot|s)ds=\int\lim_{K\rightarrow\infty}\sum_{k=1}^{K}a_{K,k}(t)w_{k}(s)F_{Y_{1}|\eta}(\cdot|s)ds\nonumber \\
 & \equiv\int\kappa(t,s)F_{Y_{1}|\eta}(\cdot|s)ds,\label{eq:RL_proof-2}
\end{align}
where $\kappa(t,s)\equiv\lim_{K\rightarrow\infty}\sum_{k=1}^{K}a_{K,k}(t)w_{k}(s)$
serves as a Dirac delta function. Because $F_{Y_{1}|\eta}(\cdot|t)=\int\kappa(t,s)F_{Y_{1}|\eta}(\cdot|s)ds$
if and only if $F_{Y_{1}|\eta}(\cdot|t)\le\int\kappa(t,s)F_{Y_{1}|\eta}(\cdot|s)ds$
and $F_{Y_{1}|\eta}(\cdot|t)\ge\int\kappa(t,s)F_{Y_{1}|\eta}(\cdot|s)ds$,
we have, by Condition \ref{condi_2}, 
\begin{equation}
F_{Y_{0}|\eta}(\cdot|t)=\int\kappa(t,s)F_{Y_{0}|\eta}(\cdot|s)ds=\lim_{K\rightarrow\infty}\sum_{k=1}^{K}a_{K,k}(t)G_{0,k}(\cdot)\label{eq:RL_proof-3}
\end{equation}
using the definition of $G_{0,k}(\cdot)$. Combining \eqref{eq:RL_proof-2},
\eqref{eq:RL_proof-3} and \eqref{eq:RL_G3}, for any $y\in\mathbb{R}$
and $t\in\mathcal{T}$, we have
\begin{align*}
F_{Y_{0}|\eta}(\psi_{\infty}(y)|t) & =\lim_{K\rightarrow\infty}\sum_{k=1}^{K}a_{K,k}(t)G_{0,k}(\psi_{\infty}(y))=\lim_{K\rightarrow\infty}\sum_{k=1}^{K}a_{K,k}(t)\lambda_{\infty}(y)G_{1,k}(y)\\
 & =\lambda_{\infty}(y)F_{Y_{1}|\eta}(y|t),
\end{align*}
which completes the proof. $\boxempty$

\subsection{Equivalence Between Rank Linearity and Condition \ref{condi_2}:
Discrete $Y_{d}$}

For $d\in\{0,1\}$, suppose $Y_{d}$ and $\eta$ are discretely distributed.
Specifically, let $\mathcal{Y}_{d}\equiv\big\{ y_{d,1},\cdots,y_{d,K_{d}}\big\}$
and $\mathcal{T}\equiv\{t_{1},\cdots,t_{K_{\eta}}\}$ be the support
of $Y_{d}$ and $\eta$, respectively. Note that even with $K_{0}=K_{1}$,
we allow that $Y_{0}$ and $Y_{1}$ have different supports (i.e.,
allowing for a ``drift''). Suppress $(Z,X)$ for simplicity.

\begin{condition}\label{condi_2-1}For arbitrary non-negative weights
$\{w_{1},\cdots,w_{K_{\eta}}\}$ and $\{\tilde{w}_{1},\cdots,\tilde{w}_{K_{\eta}}\}$
such that $\sum_{k=1}^{K_{\eta}}w_{k}=1$ and $\sum_{k=1}^{K_{\eta}}\tilde{w}_{k}=1$,
it holds that 
\[
\sum_{k=1}^{K_{\eta}}w_{k}F_{Y_{1}|\eta}(\cdot|t_{k})\leq\sum_{k=1}^{K_{\eta}}\tilde{w}_{k}F_{Y_{1}|\eta}(\cdot|t_{k})
\]
if and only if 
\[
\sum_{k=1}^{K_{\eta}}w_{k}F_{Y_{0}|\eta}(\cdot|t_{k})\leq\sum_{k=1}^{K_{\eta}}\tilde{w}_{k}F_{Y_{0}|\eta}(\cdot|t_{k}).
\]
\end{condition}

This condition can be motivated by the discussion in Remark \ref{rem:compliance_type}.

\begin{theorem}\label{thm:equiv_RL-1}For any probability distribution
function $\tilde{F}_{d}$ supported on $\mathcal{Y}_{d}\equiv\{y_{d,1},\cdots,y_{d,K_{d}}\}$,
suppose there always exists a sequence $\{c_{d,1},\cdots,c_{d,K_{\eta}}\}$
such that 
\begin{equation}
\tilde{F}_{d}(\cdot)=\sum_{k=1}^{k_{\eta}}c_{d,k}F_{Y_{d}|\eta}(\cdot|t_{k}),\label{eq:rank_condi-1}
\end{equation}
Then, Condition \ref{condi_2-1} holds if and only if (i) $K_{0}=K_{1}$
and (ii) for some strictly increasing mapping $\psi:\{y_{0,1},\cdots,y_{0,K_{0}}\}\rightarrow\{y_{1,1},\cdots,y_{1,K_{1}}\}$
and some $\lambda:\{y_{0,1},\cdots,y_{0,K_{0}}\}\rightarrow\mathbb{R}_{+}$,
\begin{equation}
F_{Y_{0}|\eta}(y|t_{k})=\lambda(y)F_{Y_{1}|\eta}(\psi(y)|t_{k}),\qquad\text{for }y\in\mathcal{Y}_{0},k=1,\cdots,K_{\eta}.\label{eq:RL}
\end{equation}

\end{theorem}

The condition \eqref{eq:rank_condi-1} is a rank condition as the
rank of matrix $\{F_{Y_{d}|\eta}(y_{d,j}|t_{j'}):j=1,...,K_{d},\quad j'=1,\cdots,k_{\eta}\}$
should be no smaller than $K_{d}$. A necessary condition is $K_{\eta}\ge K_{d}$,
namely, the support of $\eta$ is no coarser than the support of $Y_{d}$.
The rank condition would be violated when there is no endogeneity
(i.e., $Y_{d}\perp\eta$), which is not our focus. Again, the rank
condition is only introduced in this theorem to establish the relationship
between rank linearity (and hence rank similarity) and the range of
identifying conditions of this paper, and it is not necessary for
our bound analysis.

\proof By Condition \ref{condi_2-1}, we have 
\begin{equation}
\sum_{k=1}^{K_{\eta}}\delta_{k}F_{Y_{1}|\eta}(\cdot|t_{k})\leq0\text{ if and only if }\sum_{k=1}^{K_{\eta}}\delta_{k}F_{Y_{0}|\eta}(\cdot|t_{k})\leq0\label{eq:RL_proof-4}
\end{equation}
for any $K_{\eta}$-dimensional vector $\delta\equiv(\delta_{1},\cdots,\delta_{n})$
that satisfies $\sum_{k=1}^{K_{\eta}}\delta_{k}=0$.

Note that $\big(F_{Y_{1}|\eta}(y|t_{1}),\cdots,(F_{Y_{1}|\eta}(y|t_{K_{\eta}})\big)$
for each $y\in\mathcal{Y}_{1}/\{y_{K_{1}}\}$ generates an extreme
ray of the $(K_{\eta}-1)$-dimensional polar cone of a cone 
\[
\left\{ \delta\in\mathbb{R}^{n}:\sum_{k=1}^{K_{\eta}}\delta_{k}F_{Y_{1}|\eta}(\cdot|t_{k})\leq0;\sum_{k=1}^{K_{\eta}}\delta_{k}=0\right\} .
\]
A similar argument holds for $\big(F_{Y_{0}|\eta}(\cdot|t_{1}),\cdots,(F_{Y_{0}|\eta}(\cdot|t_{K_{\eta}})\big)$.
By \eqref{eq:RL_proof-4}, these two polar cones are the same. Therefore,
for each $y_{k}\in\mathcal{Y}_{1}/\{y_{K_{1}}\}$, there exists a
$y_{k}^{*}\in\mathcal{Y}_{0}/\{y_{K_{0}}\}$ such that 
\[
\big(F_{Y_{0}|\eta}(y_{k}^{*}|t_{1}),\cdots,F_{Y_{0}|\eta}(y_{k}^{*}|t_{K_{0}}))\big)=\lambda(y_{k})\times\big(F_{Y_{1}|\eta}(y_{k}|t_{1}),\cdots,F_{Y_{1}|\eta}(y_{k}|t_{K_{1}}))\big).
\]
Finally it is easy to show that if $y_{k}<y_{k'}$ then $y_{k}^{*}<y_{k'}^{*}$
and thus $\psi(y_{k})=y_{k}^{*}$ is a strictly increasing function.

\subsection{Proof of Theorem \ref{thm:main_point_ID}}

We suppress $X$ for simplicity. The proof is analogous to that of
Theorem \ref{thm:main}. Using the same notation as the earlier proof,
\eqref{eq:counter_k_point_ID} can be rewritten as
\begin{multline*}
\sum_{\ell=1}^{\ell^{*}}\frac{q(z_{\ell})-\gamma_{\ell}p(z_{\ell})}{a}\times P[Y_{1}\leq y|D=1,Z=z_{\ell}]+\sum_{\ell=\ell^{*}+1}^{L}\frac{q(z_{\ell})}{a}\times P[Y_{1}\leq y|D=1,Z=z_{\ell}]\\
=\sum_{\ell=\ell^{*}+1}^{L}\frac{\gamma_{\ell}p(z_{\ell})}{a}\times P[Y_{1}\leq y|D=1,Z=z_{\ell}].
\end{multline*}
The above equation being satisfied as equality can be viewed as being
satisfied as inequalities ``$\le$'' and ``$\ge$.'' Therefore,
by Condition \ref{as:condi_1} applied for both inequalities, we have
\begin{multline*}
\sum_{\ell=1}^{k}\frac{q(z_{\ell})-\gamma_{\ell}p(z_{\ell})}{a}\times P[Y_{0}\leq y|D=1,Z=z_{\ell}]+\sum_{\ell=\ell^{*}+1}^{L}\frac{q(z_{\ell})}{a}\times P[Y_{0}\leq y|D=1,Z=z_{\ell}]\\
=\sum_{\ell=\ell^{*}+1}^{L}\frac{\gamma_{\ell}p(z_{\ell})}{a}\times P[Y_{0}\leq y|D=1,Z=z_{\ell}].
\end{multline*}
Equivalently, we have
\begin{align*}
P[Y_{0}\leq y|D=1] & =P[Y_{0}\leq y]\times\sum_{\ell=1}^{L}\gamma_{\ell}-\sum_{\ell=1}^{L}\gamma_{\ell}\times P[Y\leq y,D=0|Z=z_{\ell}]\\
 & =-\sum_{\ell=1}^{L}\gamma_{\ell}\times P[Y\leq y,D=0|Z=z_{\ell}]
\end{align*}
by $\sum_{\ell=1}^{L}\gamma_{\ell}=0$. $\square$

\subsection{Proof of Lemma \ref{lem:models23}}

Part (i) can be shown analogous to the proof of Theorem \ref{thm:model1}.
Suppose 
\[
\int w(t,x)F_{Y_{1}|\eta,X}(\cdot|t,x)dt\leq\int\tilde{w}(t,x)F_{Y_{1}|\eta,X}(\cdot|t,x)dt
\]
holds for some $w$ and $\tilde{w}$. We want to show that 
\[
\int w(t,x)F_{Y_{0}|\eta,X}(\cdot|t,x)dt\leq\int\tilde{w}(t,x)F_{Y_{0}|\eta,X}(\cdot|t,x)dt.
\]
First, because of the strict monotonicity of $q(d,x,\cdot)$, we have
\[
\int w(t,x)F_{U_{1}|\eta,X}(\cdot|t,x)dt\leq\int\tilde{w}(t,x)F_{U_{1}|\eta,X}(\cdot|t,x)dt
\]
and it suffices to show 
\[
\int w(t,x)F_{U_{0}|\eta,X}(\cdot|t,x)dt\leq\int\tilde{w}(t,x)F_{U_{0}|\eta,X}(\cdot|t,x)dt.
\]
Second, for any $v\in\text{Supp}(V|X=x)$, because of the strict
monotonicity of $\phi(\cdot,v)$, we have $1(U_{1}\leq u_{1})\overset{a.s.}{=}1(\phi(U_{1},v)\leq\phi(u_{1},v))$.
Because $V\bot(U_{1},\eta)|X$, we have 
\[
\int w(t,x)F_{\phi(U_{1},V)|\eta,X,V}(\phi(\cdot,v)|t,x,v)dt\leq\int\tilde{w}(t,x)F_{\phi(U_{1},V)|\eta,X,V}(\phi(\cdot,v)|t,x,v)dt
\]
and thus, 
\[
\int w(t,x)F_{U_{0}|\eta,X,V}(\phi(\cdot,v)|t,x,v)dt\leq\int\tilde{w}(t,x)F_{U_{0}|\eta,X,V}(\phi(\cdot,v)|t,x,v)dt.
\]
Conditional on $(\eta,X,V)$, $\text{Supp}(\phi(U_{1},v))=\text{Supp}(\phi(U_{1},V))=\text{Supp}(U_{0})$.
Therefore, for $u_{0}$ in that support, 
\[
\int w(t,x)F_{U_{0}|\eta,X,V}(u_{0}|t,x,v)dt\leq\int\tilde{w}(t,x)F_{U_{0}|\eta,X,V}(u_{0}|t,x,v)dt.
\]
It follows that
\begin{align*}
 & \int\int w(t,x)F_{U_{0}|\eta,X,V}(u_{0}|t,x,v)f_{V|X}(v|x)dtdv\\
\leq & \int\int\tilde{w}(t,x)F_{U_{0}|\eta,X,V}(u_{0}|t,x,v)f_{V|X}(v|x)dtdv
\end{align*}
 Note that $f_{V|X}=f_{V|\eta,X}$. Then, by the law of iterated expectation,
we have 
\[
\int w(t,x)F_{U_{0}|\eta,X}(u_{0}|t,x)dt\leq\int\tilde{w}(t,x)F_{U_{0}|\eta,X}(u_{0}|t,x)dt.
\]

Next, we prove part (ii) by first noting that
\begin{align*}
\Pr[U_{0}\le u|\eta,X] & =\Pr[\phi(U_{1})\le u,V\le u|\eta,X]=\Pr[\phi(U_{1})\le u|\eta,X]\Pr[V\le u|X].
\end{align*}
Therefore,
\begin{align*}
F_{Y_{0}|\eta,X}(y|t,x) & =\Pr[g(0,x,U_{0})\le y|\eta=t,X=x]=\Pr[U_{0}\le g^{-1}(0,x,y)|\eta=t,X=x]\\
 & =\Pr[\phi(U_{1})\le g^{-1}(0,x,y)|\eta=t,X=x]\Pr[V\le g^{-1}(0,x,y)]\\
 & =\Pr[Y_{1}\le g(1,x,\phi^{-1}(g^{-1}(0,x,y)))|\eta=t,X=x]\Pr[V\le g^{-1}(0,x,y)]\\
 & =F_{Y_{1}|\eta,X}(\psi(y,x)|t,x)\lambda(y,x),
\end{align*}
where $\psi(y,x)\equiv g(1,x,\phi^{-1}(g^{-1}(0,x,y)))$ and $\lambda(y,x)\equiv F_{V}(g^{-1}(0,x,y))$.
$\square$

\subsection{Proof of Theorem \ref{thm:strong_duality}}

The proof is immediate by applying Theorem 6.9 in \citet{hettich1993semi}.
This is because (i) the primal problem is superconsistent as both
$\boldsymbol{p}(y,1)$ and $p(y|1)$ are continuous on compact $\mathcal{Y}$
and (ii) $\gamma^{*}\in\{y:\boldsymbol{p}(y,1)'\gamma\ge p(y|1)\}$
such that $\boldsymbol{p}(y,1)'\gamma^{*}>p(y|1)$. $\square$

\end{appendix}

\bibliographystyle{ecta}
\bibliography{multi_IVs}

\begin{thebibliography}{32}
\newcommand{\enquote}[1]{``#1''}
\expandafter\ifx\csname natexlab\endcsname\relax\def\natexlab#1{#1}\fi

\bibitem[\protect\citeauthoryear{Abadie, Angrist, and Imbens}{Abadie
  et~al.}{2002}]{abadie2002instrumental}
\textsc{Abadie, A., J.~Angrist, and G.~Imbens} (2002): \enquote{Instrumental
  variables estimates of the effect of subsidized training on the quantiles of
  trainee earnings,} \emph{Econometrica}, 70, 91--117.

\bibitem[\protect\citeauthoryear{Blundell, Gosling, Ichimura, and
  Meghir}{Blundell et~al.}{2007}]{blundell2007changes}
\textsc{Blundell, R., A.~Gosling, H.~Ichimura, and C.~Meghir} (2007):
  \enquote{Changes in the distribution of male and female wages accounting for
  employment composition using bounds,} \emph{Econometrica}, 75, 323--363.

\bibitem[\protect\citeauthoryear{Calafiore and Campi}{Calafiore and
  Campi}{2005}]{calafiore2005uncertain}
\textsc{Calafiore, G. and M.~C. Campi} (2005): \enquote{Uncertain convex
  programs: randomized solutions and confidence levels,} \emph{Mathematical
  Programming}, 102, 25--46.

\bibitem[\protect\citeauthoryear{Chernozhukov and Hansen}{Chernozhukov and
  Hansen}{2005}]{chernozhukov2005iv}
\textsc{Chernozhukov, V. and C.~Hansen} (2005): \enquote{An {IV} model of
  quantile treatment effects,} \emph{Econometrica}, 73, 245--261.

\bibitem[\protect\citeauthoryear{Chernozhukov and Hansen}{Chernozhukov and
  Hansen}{2013}]{chernozhukov2013quantile}
---\hspace{-.1pt}---\hspace{-.1pt}--- (2013): \enquote{Quantile models with
  endogeneity,} \emph{Annu. Rev. Econ.}, 5, 57--81.

\bibitem[\protect\citeauthoryear{Chesher}{Chesher}{2003}]{chesher2003identification}
\textsc{Chesher, A.} (2003): \enquote{Identification in nonseparable models,}
  \emph{Econometrica}, 71, 1405--1441.

\bibitem[\protect\citeauthoryear{Chesher}{Chesher}{2005}]{Che05}
---\hspace{-.1pt}---\hspace{-.1pt}--- (2005): \enquote{Nonparametric
  identification under discrete variation,} \emph{Econometrica}, 73,
  1525--1550.

\bibitem[\protect\citeauthoryear{D'Haultf{\oe}uille and
  F{\'e}vrier}{D'Haultf{\oe}uille and F{\'e}vrier}{2015}]{d2015identification}
\textsc{D'Haultf{\oe}uille, X. and P.~F{\'e}vrier} (2015):
  \enquote{Identification of nonseparable triangular models with discrete
  instruments,} \emph{Econometrica}, 83, 1199--1210.

\bibitem[\protect\citeauthoryear{Dong and Shen}{Dong and
  Shen}{2018}]{dong2018testing}
\textsc{Dong, Y. and S.~Shen} (2018): \enquote{Testing for rank invariance or
  similarity in program evaluation,} \emph{Review of Economics and Statistics},
  100, 78--85.

\bibitem[\protect\citeauthoryear{Frandsen and Lefgren}{Frandsen and
  Lefgren}{2018}]{frandsen2018testing}
\textsc{Frandsen, B.~R. and L.~J. Lefgren} (2018): \enquote{Testing rank
  similarity,} \emph{Review of Economics and Statistics}, 100, 86--91.

\bibitem[\protect\citeauthoryear{Han}{Han}{2021}]{han2021identification}
\textsc{Han, S.} (2021): \enquote{Identification in nonparametric models for
  dynamic treatment effects,} \emph{Journal of Econometrics}, 225, 132--147.

\bibitem[\protect\citeauthoryear{Han and Yang}{Han and
  Yang}{2023}]{han2020sharp}
\textsc{Han, S. and S.~Yang} (2023): \enquote{A Computational Approach to
  Identification of Treatment Effects for Policy Evaluation,} \emph{arXiv
  preprint arXiv:2009.13861}.

\bibitem[\protect\citeauthoryear{Heckman}{Heckman}{1990}]{heckman1990varieties}
\textsc{Heckman, J.} (1990): \enquote{Varieties of selection bias,} \emph{The
  American Economic Review}, 80, 313--318.

\bibitem[\protect\citeauthoryear{Heckman, Smith, and Clements}{Heckman
  et~al.}{1997}]{heckman1997making}
\textsc{Heckman, J.~J., J.~Smith, and N.~Clements} (1997): \enquote{Making the
  most out of programme evaluations and social experiments: Accounting for
  heterogeneity in programme impacts,} \emph{The Review of Economic Studies},
  64, 487--535.

\bibitem[\protect\citeauthoryear{Hettich and Kortanek}{Hettich and
  Kortanek}{1993}]{hettich1993semi}
\textsc{Hettich, R. and K.~O. Kortanek} (1993): \enquote{Semi-infinite
  programming: theory, methods, and applications,} \emph{SIAM review}, 35,
  380--429.

\bibitem[\protect\citeauthoryear{Imbens and Angrist}{Imbens and
  Angrist}{1994}]{imbens1994identification}
\textsc{Imbens, G.~W. and J.~D. Angrist} (1994): \enquote{Identification and
  Estimation of Local Average Treatment Effects,} \emph{Econometrica}, 62,
  467--475.

\bibitem[\protect\citeauthoryear{Jun, Pinkse, and Xu}{Jun
  et~al.}{2011}]{jun2011tighter}
\textsc{Jun, S.~J., J.~Pinkse, and H.~Xu} (2011): \enquote{Tighter bounds in
  triangular systems,} \emph{Journal of Econometrics}, 161, 122--128.

\bibitem[\protect\citeauthoryear{Kim and Park}{Kim and
  Park}{2022}]{kim2022testing}
\textsc{Kim, J.~H. and B.~G. Park} (2022): \enquote{Testing rank similarity in
  the local average treatment effects model,} \emph{Econometric Reviews},
  1--22.

\bibitem[\protect\citeauthoryear{Maasoumi and Wang}{Maasoumi and
  Wang}{2019}]{maasoumi2019gender}
\textsc{Maasoumi, E. and L.~Wang} (2019): \enquote{The gender gap between
  earnings distributions,} \emph{Journal of Political Economy}, 127,
  2438--2504.

\bibitem[\protect\citeauthoryear{Manski}{Manski}{1990}]{manski1990nonparametric}
\textsc{Manski, C.~F.} (1990): \enquote{Nonparametric bounds on treatment
  effects,} \emph{The American Economic Review}, 80, 319--323.

\bibitem[\protect\citeauthoryear{Manski}{Manski}{1994}]{manski1994selection}
---\hspace{-.1pt}---\hspace{-.1pt}--- (1994): \enquote{The selection problem,}
  in \emph{Advances in Econometrics, Sixth World Congress, ed. by C. Sims},
  vol.~1, 143--70.

\bibitem[\protect\citeauthoryear{Manski}{Manski}{1997}]{manski1997monotone}
---\hspace{-.1pt}---\hspace{-.1pt}--- (1997): \enquote{Monotone treatment
  response,} \emph{Econometrica: Journal of the Econometric Society},
  1311--1334.

\bibitem[\protect\citeauthoryear{Manski and Pepper}{Manski and
  Pepper}{2000}]{MP00}
\textsc{Manski, C.~F. and J.~V. Pepper} (2000): \enquote{Monotone instrumental
  variables: With an application to the returns to schooling,}
  \emph{Econometrica}, 68, 997--1010.

\bibitem[\protect\citeauthoryear{Mogstad, Santos, and Torgovitsky}{Mogstad
  et~al.}{2018}]{mogstad2018using}
\textsc{Mogstad, M., A.~Santos, and A.~Torgovitsky} (2018): \enquote{Using
  instrumental variables for inference about policy relevant treatment
  parameters,} \emph{Econometrica}, 86, 1589--1619.

\bibitem[\protect\citeauthoryear{Mogstad, Torgovitsky, and Walters}{Mogstad
  et~al.}{2021}]{mogstad2021causal}
\textsc{Mogstad, M., A.~Torgovitsky, and C.~R. Walters} (2021): \enquote{The
  causal interpretation of two-stage least squares with multiple instrumental
  variables,} \emph{American Economic Review}, 111, 3663--98.

\bibitem[\protect\citeauthoryear{Pomatto, Strack, and Tamuz}{Pomatto
  et~al.}{2020}]{pomatto2020stochastic}
\textsc{Pomatto, L., P.~Strack, and O.~Tamuz} (2020): \enquote{Stochastic
  dominance under independent noise,} \emph{Journal of Political Economy}, 128,
  1877--1900.

\bibitem[\protect\citeauthoryear{Rubin}{Rubin}{1974}]{rubin1974estimating}
\textsc{Rubin, D.~B.} (1974): \enquote{Estimating causal effects of treatments
  in randomized and nonrandomized studies.} \emph{Journal of Educational
  Psychology}, 66, 688.

\bibitem[\protect\citeauthoryear{Shaikh and Vytlacil}{Shaikh and
  Vytlacil}{2011}]{SV11}
\textsc{Shaikh, A.~M. and E.~J. Vytlacil} (2011): \enquote{Partial
  identification in triangular systems of equations with binary dependent
  variables,} \emph{Econometrica}, 79, 949--955.

\bibitem[\protect\citeauthoryear{Torgovitsky}{Torgovitsky}{2015}]{torgovitsky2015identification}
\textsc{Torgovitsky, A.} (2015): \enquote{Identification of nonseparable models
  using instruments with small support,} \emph{Econometrica}, 83, 1185--1197.

\bibitem[\protect\citeauthoryear{Vuong and Xu}{Vuong and
  Xu}{2017}]{vuong2017counterfactual}
\textsc{Vuong, Q. and H.~Xu} (2017): \enquote{Counterfactual mapping and
  individual treatment effects in nonseparable models with binary endogeneity,}
  \emph{Quantitative Economics}, 8, 589--610.

\bibitem[\protect\citeauthoryear{Vytlacil}{Vytlacil}{2002}]{vytlacil2002independence}
\textsc{Vytlacil, E.} (2002): \enquote{Independence, monotonicity, and latent
  index models: An equivalence result,} \emph{Econometrica}, 70, 331--341.

\bibitem[\protect\citeauthoryear{Vytlacil and Yildiz}{Vytlacil and
  Yildiz}{2007}]{VY07}
\textsc{Vytlacil, E. and N.~Yildiz} (2007): \enquote{Dummy endogenous variables
  in weakly separable models,} \emph{Econometrica}, 75, 757--779.

\end{thebibliography}

\end{document}